\documentclass[structabstract]{aa}  
\newcommand{\bq}{\begin{eqnarray}}
\newcommand{\eq}{\end{eqnarray}}                                   
\newcommand{\teff}{T_{\mathrm{eff}}}
\newcommand{\mD}{\ensuremath{\left\langle\mathrm{3D}\right\rangle}}
\usepackage{graphicx,natbib}
\usepackage{txfonts}
\begin{document}
   \title{3D simulations of M star atmosphere velocities and their influence
   on molecular \element[][]{FeH} lines }

   \subtitle{}

   \author{S.Wende
          \inst{1}
          \and
          A.Reiners
	  \inst{1}
	  \and
	  H.-G.Ludwig
	  \inst{2}
          }

   \institute{Institut f\"ur Astrophysik, Georg-August-Universit\"at
  G\"ottingen, Friedrich-Hund Platz 1, D-37077, Germany\\
              \email{sewende@astro.physik.uni-goettingen.de}
	      \email{Ansgar.Reiners@phys.uni-goettingen.de}
         \and
             GEPI, CIFIST, Observatoire de Paris-Meudon, 5 place Jules Janssen,
  92195 Meudon Cedex, France\\
             \email{Hans.Ludwig@obspm.fr}
             }

   \date{Received 19 August 2009 / Accepted 13 October 2009}

 
  \abstract 
 {The measurement of line broadening in cool stars is in general a difficult
  task. In order to detect slow rotation or weak magnetic fields, an accuracy
  of $1$\,km\,s$^{-1}$ is needed. In this regime the broadening from
  convective motion become important. We present an investigation of the
  velocity fields in early to late M-type star hydrodynamic models, and we simulate
  their influence on \element[][]{FeH} molecular line shapes. The M star
  model parameters range between $\log{g}$ of $3.0~-~5.0$ and effective
  temperatures of $2500$\,K and $4000$\,K.}  
{Our aim is to characterize the $\teff$- and $\log{g}$-dependence of the
  velocity fields and express them in terms of micro- and macro-turbulent
  velocities in the one dimensional sense. We present also a direct comparison
  between 3D hydrodynamical velocity fields and 1D turbulent
  velocities. The velocity fields strongly affect the line shapes of
  \element[][]{FeH}, and it is our goal to give a rough estimate for the $\log{g}$
  and $\teff$ parameter range in which 3D spectral synthesis is necessary and
  where 1D synthesis suffices. Eventually we want to distinguish
  between the velocity-broadening from convective motion and the rotational- or
  Zeeman-broadening in M-type stars which we are planning to measure. 
  For the latter \element[][]{FeH} lines are an important indicator.}  
{In order to calculate M-star structure models we employ the 3D
  radiative-hydrodynamics (RHD) code \texttt{CO$^5$BOLD}. The
  spectral synthesis on these models is performed with the line synthesis code
  \texttt{LINFOR3D}. We describe the 3D velocity fields in terms of a Gaussian
  standard deviation and project them onto the line of sight to include
  geometrical and limb-darkening effects. The micro- and macro-turbulent
  velocities are determined with the ``Curve of Growth'' method and convolution
  with a Gaussian velocity profile, respectively. To characterize the $\log{g}$
  and $\teff$ dependence of \element[][]{FeH} lines, the equivalent width,
  line width, and line depth are regarded.} 
{The velocity fields in M-stars strongly depend on  $\log{g}$ and
  $\teff$. They become stronger with decreasing $\log{g}$ and
  increasing $\teff$. The projected velocities from the 3D models agree
  within $\sim~100$\,m\,s$^{-1}$ with the 1D micro- and macro-turbulent
  velocities. The \element[][]{FeH} 
  line quantities systematically depend on $\log{g}$ and $\teff$.}  
{The influence of hydrodynamical velocity fields on line shapes of M-type
  stars can well be reproduced with 1D broadening methods.  \element[][]{FeH}
  lines turn out to provide a mean to measure $\log{g}$ and $\teff$ in M-type
  stars.  Since different \element[][]{FeH} lines behave all in a similar
  manner, they provide an ideal measure for rotational and magnetic
  broadening. }


   \maketitle
%

\section{Introduction}
Most of our knowledge about stars comes from spectroscopic
investigation of atomic or molecular lines. In sun-like and hotter
stars, the strength and shape of atomic spectral lines provides
information on atmospheric structure, velocity fields, rotation,
magnetic fields, etc. Measuring the effects of velocity fields on the
shape of spectral lines requires a spectral resolving power between $R
\sim 10,000$ ($\Delta v = 30$\,km\,s$^{-1}$) for rapid stellar rotation,
$R \ga 30,000$ ($\Delta v = 10$\,km\,s$^{-1}$) for slower rotation and
high turbulent velocities, and resolution on the order of $R \sim 100,000$
for the analysis of Zeeman splitting and line shape variations due to slow
convective motion.

In slowly rotating sun-like stars, usually a large number of relatively
isolated spectral lines are available for the investigation of Doppler
broadened spectral lines. These lines are embedded in a clearly visible
continuum allowing a detailed analysis of individual lines at high
precision. At cooler temperature, first the number of atomic lines is
increasing so that more and more lines become blended rendering the
investigation of individual lines more difficult.  At temperatures around
4000\,K, molecular lines, predominantly \element[][]{VO} and
\element[][]{TiO}, start to become important. At optical wavelengths,
molecular bands in general consist of many lines that are blended so that the
absorption mainly appears as an absorption band; individual molecular lines
are difficult to identify. At temperatures in the M type stars regime (4000\,K
and less), atomic lines start to vanish because atoms are mainly neutral and
higher ionization levels are weakly populated. Only alkali lines appear that
are strongly affected by pressure broadening.  Thus, the detailed
spectroscopic investigation of velocity fields in M dwarfs is very difficult
at optical wavelengths.

M-type stars emit the bulk of their flux at infrared wavelengths redward of
1\,$\mu$m. This implies that observation of high SNR spectra in principle is
easier in the infrared. Furthermore, M type stars exhibit a number of
molecular absorption bands in the infrared, for example \element[][]{FeH}.  In
these bands, the individual lines are relatively well separated and provide a
good tracer of stellar velocity fields. The lines are intrinsically much
narrower than atomic lines in sun-like stars because Doppler broadening due to
the temperature related motion of the atoms and molecules is much
reduced. Thus, the lines can be used for the whole arsenal of line profile
analysis that has been applied successfully to sun-like stars over the last
decades.

Examples of analyses using \element[][]{FeH} lines are the investigation of the
rotation activity connection in field M-dwarfs, which requires the
measurement of rotational line broadening with an accuracy of $1$\,km\,s$^{-1}$
\citep{2007A&A...467..259R}. Another example is the measurement of
magnetic fields comparing Zeeman broadening in magnetically sensitive
and insensitive absorption lines \citep[see
e.g.~][]{2006ApJ...644..497R}. A precise analysis of \element[][]{FeH} lines,
however is only
possible if the underlying velocity fields of the M dwarfs atmospheres
are thoroughly understood. In this paper, we model the surface
velocity fields of M type stars and their influence on the narrow spectral
lines of \element[][]{FeH}.

We calculate 3D-\texttt{CO$^5$BOLD} structure models
\citep{2002A&A...395...99L} which serve as an input for the line
formation program \texttt{LINFOR3D} \citep[based on][]{Bascheck1966}.
Turbulence's are included in a natural way using hydrodynamics, so
that we are able to investigate the modeled spectral lines for effects
from micro- and macro-turbulent velocities in the classical sense and
their influence on the line shapes. The comparison with 1D-models
gives a rough estimate of the necessity of using 3D-models in the
spectral domain of cool stars. In the first part of this paper we
investigate the velocity fields in the models and their dependence on
$\log{g}$ and $\teff$. In the second part, we investigate the
influence of velocity fields, $\log{g}$, and $\teff$ on the
\element[][]{FeH} molecular lines.

\section{3D model atmospheres}

The three-dimensional time-dependent model atmospheres (hereafter ``3D
models'') are calculated with the radiation-hydrodynamics code
\texttt{CO$^5$BOLD} (abbreviation for ``COnservative COde for the COmputation
of COmpressible COnvection in a BOx of L Dimensions with L=2,3''). It is
designed to model solar and stellar surface convection. For solar-like stars
like the M-type objects considered here, \texttt{CO$^5$BOLD} employs a local
set-up in which the governing equations are solved in a small (relative to the
stellar radius) Cartesian domain located at the stellar surface (``box in a
star set-up''). The optically thin stellar photosphere and the upper-most part
of the underlying convective envelope are embedded in the computational domain.
\texttt{CO$^5$BOLD} solves the coupled non-linear equations of compressible
hydrodynamics in an external gravitational field in three spatial dimensions
\citep{2002AN....323..213F,2004A&A...414.1121W} together with non-local
frequency-dependent radiative transfer.  In these 3D models
convection is treated without any assumptions like in 1D mixing-length theory.
The velocity fields and its related transport properties are a direct result
of the solution of the hydrodynamic equations. Due to this,
\texttt{CO$^5$BOLD} is a well-suited tool to investigate the influence of
velocity fields on spectral line shapes. A \texttt{CO$^5$BOLD} model consists
of a sequence of 3D flow fields (``snapshots'') representing the temporal
evolution and spatial structure of the flow. To perform spectral synthesis
calculations based on the 3D \texttt{CO$^5$BOLD}-models we use the 3D line
formation code \texttt{Linfor3D}. It takes into account the full 3D thermal
structure and velocity field in the calculation of the line profiles. It
assumes strict Local Thermodynamic Equilibrium (LTE). In this paper we will
call the spectral lines computed from three-dimensional atmosphere models
``3D-lines''.

In order to analyze the influence of velocity fields in M-stars on
\element[][]{FeH} lines, we construct a set of \texttt{CO$^5$BOLD}-models with
$\teff=2500$\,K -- $4000$\,K and $\log{g}=3.0-5.0$\,[cgs]. Table \ref{tab1}
give the model parameters. In the $\teff$-sequence, we simulated
main sequence stars and varied the surface gravity slightly with increasing
effective temperature. For the $\log{g}$-sequence, we computed models with
different $\log{g}$ values aiming at the same effective temperature of
$3300$\,K but the models settle to slightly higher or lower $\teff$ values. We
decided not to adjust these resulting effective temperatures, because slight
differences in $\teff$ do not change the line profiles significantly. We
accepted the $\teff$ deviations to
avoid the large computational effort which would be necessary to adjust the
models to a common effective temperature. However,
we apply corrections to the line shape related quantities such as equivalent width
(see Section~\ref{sec:lineshapes}).

The opacities used in the \texttt{CO$^5$BOLD} model calculations originate
from the \texttt{PHOENIX} stellar atmosphere package
\citep{1999JCoAM.109...41H} assuming a solar chemical composition according to
\citet{2005ASPC..336...25A}. The opacity tables were computed after
\citet{jwfOpac05} and \citet{bdCO5BOLD}. These opacities are particularly
well-suited for our investigation since they are adapted to very cool stellar
atmospheres. The raw data consist of opacities sampled at 62,890 wavelength
points for a grid of temperatures and gas pressures. For representing the
wavelength dependence of the radiation field in the \texttt{CO$^5$BOLD} models
the opacities are re-sampled into six wavelength groups using the opacity
binning method
\citep{1982A&A...107....1N,1992HGLPhDT,1994A&A...284..105L}.
In this approach, the frequencies which reach monochromatic optical depth
unity within a certain depth range of the model atmosphere, will
be grouped into one frequency bin on the basis of their opacities.
For each investigated atmospheric parameter combination the sorting of the
wavelengths into groups is performed according the run of monochromatic
optical depth in a corresponding \texttt{PHOENIX} 1D model atmosphere. The
thresholds for the sorting are chosen in logarithmic Rosseland optical depth
as $\{+\infty, 0.0, -1.0, -2.0, -3.0, -4.5, -\infty\}$. In each group a
switching is done from a Rosseland average in the optically thick regime to a
Planck average in the optically thin regime, except for the group representing
the largest opacities, where the Rosseland average is used throughout. In this
last bin, which describes the optically thick regions, only the Rosseland average is
used because the radiative transfer
is local and can be described as a diffusive process \citep{2004A&A...421..741V}.

The horizontal size of the models provide sufficient space to allow the
development of a small number (10--20) of convective cells.  Their number has
to be large enough to avoid box-size dependent effects, but also small
enough that there is a sufficient number of grid points available to resolve
each individual cell. The size of the convective cells scales roughly 
inversely proportional to the surface gravity. Accordingly, the horizontal size
of the computational box is set to larger sizes towards lower $\log{g}$ values.
The horizontal size of the model with $\teff=3275$\,K is just large enough to
fulfill the criteria of the minimal number of 10 convective cells (see
Fig.~\ref{loggcontour}), and we saw in
test simulations that the results will not change with a larger model (in
horizontal size). Therefore we will use this well evolved model as well.
The vertical dimension is set to embed the optically thin photosphere, and a
number of pressure scale heights of the sub-photospheric layers below. We
deliberately 
keep the depth of our models rather small to avoid problems due to
numerical instabilities analogous to the ones encountered and discussed in our
previous works on the hydrodynamics of M-type stellar atmospheres
\citep{2002A&A...395...99L,2006A&A...459..599L}.

For the comparison with 1D models, we spatially average the 3D-model over
surfaces of equal Rosseland optical depth at selected instants in time. We
call the obtained sequence of 1D structures \mD-model. We follow the procedure
of \citet{1995A&A...300..473S} and average the fourth moment of the temperature
and first moment of the gas pressure to preserve the radiative properties of
the 3D-model as far as possible. The 3D velocity information is ignored in the
\mD-model and replaced by a micro- and macro-turbulent velocity.  By
construction, the \mD-model has the same thermal profile as the 3D-model, but
evidently without the horizontal inhomogeneities related to the convective
granulation pattern. We will call the spectral lines synthesized from
\mD-models ``\mD-lines''.
\begin{table*}
\centering
\caption{Overview of different model quantities for models at constant $\teff$ 
  and different $\log{g}$ (upper part) and at constant
  $\log{g}$ and different $\teff$ (lower part).} 
\begin{tabular}{lrrrrrr}
\hline\hline
 Model code & Size(x,y,z) [km] &Grid points (nx,ny,nz) & $H_p$ [km]$^{a}$& z-size [$H_p$]&$\teff$ [K] & $\log{g}$ [cgs]\\
\hline
d3t33g30mm00w1  & 85000 x 85000 x 58350 & 180 x 180 x 150  & 2821 &20.7& 3240 & 3.0 \\
d3t33g35mm00w1  & 28000 x 28000 x 11500 & 180 x 180 x 150  & 826 &13.9& 3270 & 3.5 \\ 
d3t33g40mm00w1  & 7750 x 7750 x 1850 & 150 x 150 x 150  & 250 & 7.4 & 3315 & 4.0 \\ 
d3t33g50mm00w1  & 300 x 300 x 260 & 180 x 180 x 150  & 18 & 14.5 & 3275 & 5.0 \\
 & & & & \\
d3t40g45mm00n01 & 4700 x 4700 x 1150 &140 x 140 x 141  & 109 & 10.6 & 4000 & 4.5 \\
d3t38g49mm00w1 & 1900 x 1900 x 420  & 140 x 140 x 150  & 36 & 11.7 &3820 & 4.9 \\ 
d3t35g50mm00w1 & 1070 x 1070 x 290  & 180 x 180 x 150  & 20 & 14.5& 3380 & 5.0 \\
d3t28g50mm00w1 & 370 x 370 x 270 & 250 x 250 x 140  & 13 &20.8& 2800 & 5.0 \\
d3t25g50mm00w1 & 240 x 240 x 170 & 250 x 250 x 120  & 12 &14.2& 2575 & 5.0 \\ 
\hline
\multicolumn{1}{l}{$^a$ at $\tau=1$}\\
\end{tabular}

\label{tab1}
\end{table*}

\subsection{Atmosphere structures}
\begin{figure}
\centering
 \includegraphics[width=0.5\textwidth,bb=35 55 580 794]{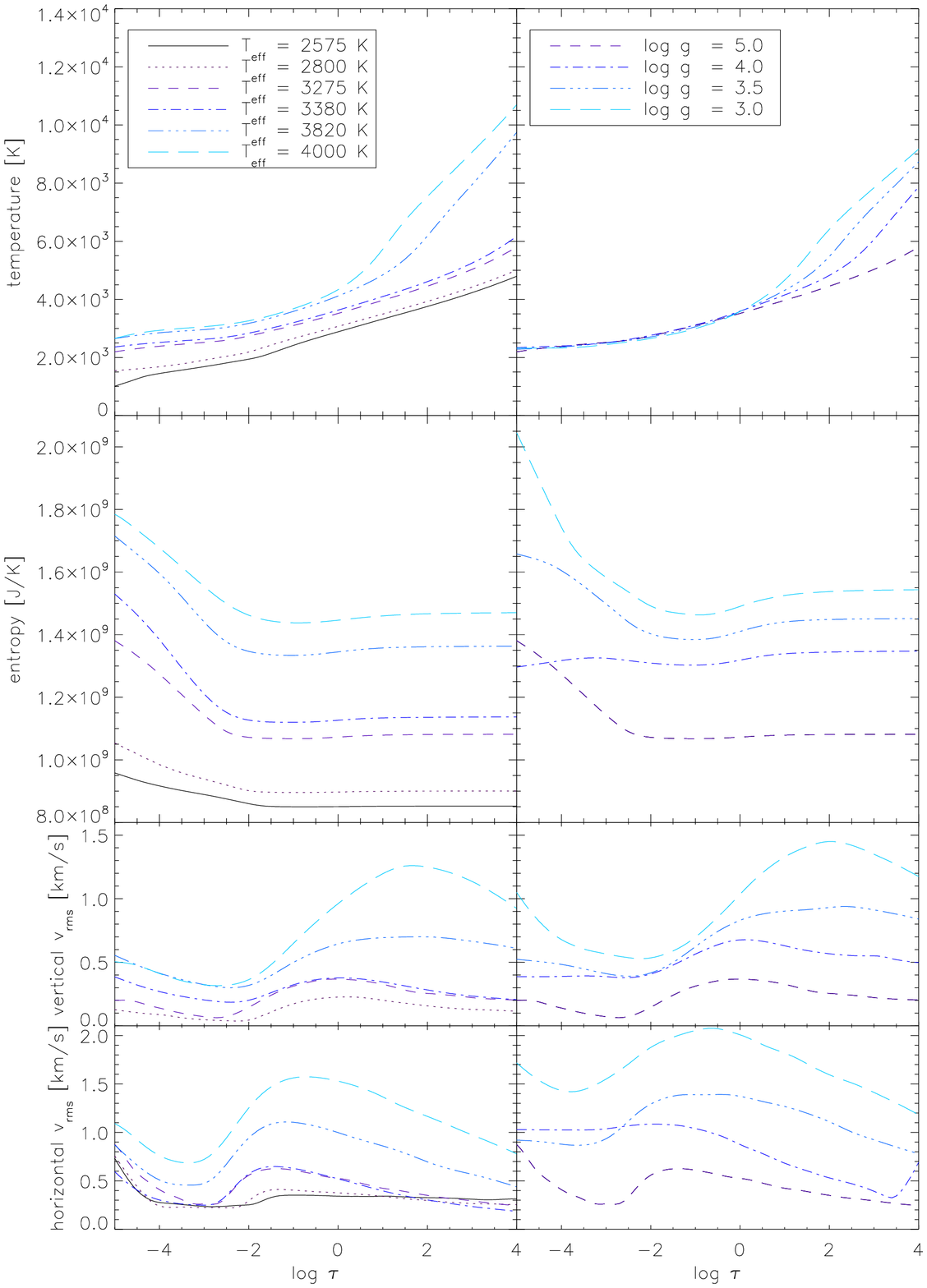}
 \caption{From top to bottom, the temperature, entropy, vertical, and horizontal velocity
 are plotted as a function of optical depth. The column on the left side shows models with 
different
 $\teff$ and constant $\log{g}$, on the right side the models are at a constant $\teff$ with 
varying $\log{g}$. The rise of horizontal velocity in the $\log{g}=4.0$\,[cgs]
 model in the deeper atmospheric layers is due to interpolation from z- to
 $\tau$-scale.}
\label{structure}
\end{figure}

The temperature stratification shown in Fig.\ref{structure} (top left) of the
models with changing $\teff$ appears very similar for all models in the region
below $\log{\tau}~\sim~1$. They reach their $\teff$ around $\tau~\sim~2/3$ and
continue to decrease to higher atmospheric layers. Above $\log{\tau~\sim~1}$
the temperature of the two hottest models increase more strongly than in the
cooler cases. Since the models are almost adiabatic in the deeper layers (see
below) the temperature gradient follows the adiabatic gradient, which is given
by the equation of state and steeper in hotter models due to the inefficient
$\element[][]{H_2}$ molecule formation. This increase of temperature is also
very similar to the lower $\log{g}$ models on the right side of
Fig.\ref{structure}. These models also show a temperature gradient which
becomes steeper to deeper atmospheric layers and with decreasing surface
gravity again due to a steeper adiabatic gradient. All models reach an almost
equal effective temperature at $\tau~\sim~2/3$ and their temperature
stratification does not differ strongly to smaller optical depth.

The entropy stratification for models with varying $\teff$ (mid left in
Fig.\ref{structure}) show a similar behavior for all models. It is adiabatic
($dS/d\tau = 0$) in layers below $\log{\tau}~\sim~1$ and has a superadiabatic
region ($dS/d\tau > 0$) between $\log{\tau}~\sim~1$ and $\sim~-1$ that moves
slightly to smaller optical depth for hotter models. In these regions, with
$dS/d\tau \geq 0$, the models are convectively unstable and become
convectively stable in the outer parts of the atmosphere where $dS/d\tau <
0$. In the models with changing $\log{g}$ (mid right in Fig.\ref{structure}),
we can see that the entropy behaves almost like in the $\teff$ case. To lower
surface gravities, the superadiabatic region is more significant. In higher
layers, the models become convectively stable except the model with
$\log{g}=4.0$\,[cgs] which shows a second decrease of entropy to the outer
part layers. To understand this behavior, we have to investigate the
adiabatic gradient of this region which is very small and changes very little
along the upper atmosphere. This is due to the equation of state used in the
models and can be seen in Fig.16 of
\citet{2006A&A...459..599L} (model H4 in this figure equates to our
$\log{g}=4.0$ model). This figure shows that the upper atmosphere lies in a plane of
small and constant adiabatic gradient. Due to this, the model becomes
convectively unstable again in the upper layers.  This is probably the reason
for the, in comparison to other models, higher velocities in this model.  

In the left bottom panel of Fig.\ref{structure}, the
horizontal and vertical rms-velocities are plotted for models with different
$\teff$. Both velocity components increase with increasing $\teff$. The maxima
of the vertical velocity moves to slightly deeper layers with higher
temperatures and the maxima of the horizontal velocity stays almost at the
same optical depth. We can see a qualitatively similar dependence in the
$\log{g}$ model sequence in the right bottom plot in Fig.\ref{structure}. Only
the model with $\log{g}=4.0$\,[cgs] shows peculiar behavior in the upper
atmospheric layers, which is probably related to the entropy stratification in
this model. We will describe the velocity fields in the models in more detail
and with a slightly different method in the next section.

\section{Velocity fields in the \texttt{CO$^5$BOLD}-models}
Before we investigate the effect of velocity fields on spectral lines, we
analyze the velocity fields in the models themselves and we will do this with
respect to spectral lines.  Spectral lines are broadened by velocity fields
where the wavelength of absorption or emission of a particle is shifted due to
its motion in the gas.  Here we are mostly concerned with the macroscopic,
hydrodynamic motions but have in mind that the thermal motions are also
constituting an important contribution. If we envision each voxel in the RHD
model cube to form its own spectral line, the whole line consists of a
(weighted) sum of single lines.  The velocity distribution might be
represented by a histogram of the velocities of the voxels which gives us the
velocity dispersion. We try to describe the velocity fields in that sense
instead of using the rms-velocities shown in Fig.\ref{structure}.  In the
\texttt{CO$^5$BOLD}-models, a velocity vector is assigned to each voxel and
consists of the velocities in x-, y-, and z-direction. We will investigate the
vertical and horizontal component of the velocity dispersion in the models and
the total velocity dispersion
$\sigma_{\mathrm{tot}}=\sqrt{\sigma_x^2+\sigma_y^2+\sigma_z^2}$. 
In order to describe the height dependent velocity dispersion we applied a binning method,
i.e. we plot all velocity components of a certain horizontal plane of equal
optical depth $\tau$ in the \texttt{CO$^5$BOLD} cube in a histogram with a bin
size of $25$\,m\,s$^{-1}$. 
\begin{figure}
\centering
 \includegraphics[width=8cm,bb=19 29 580 410]{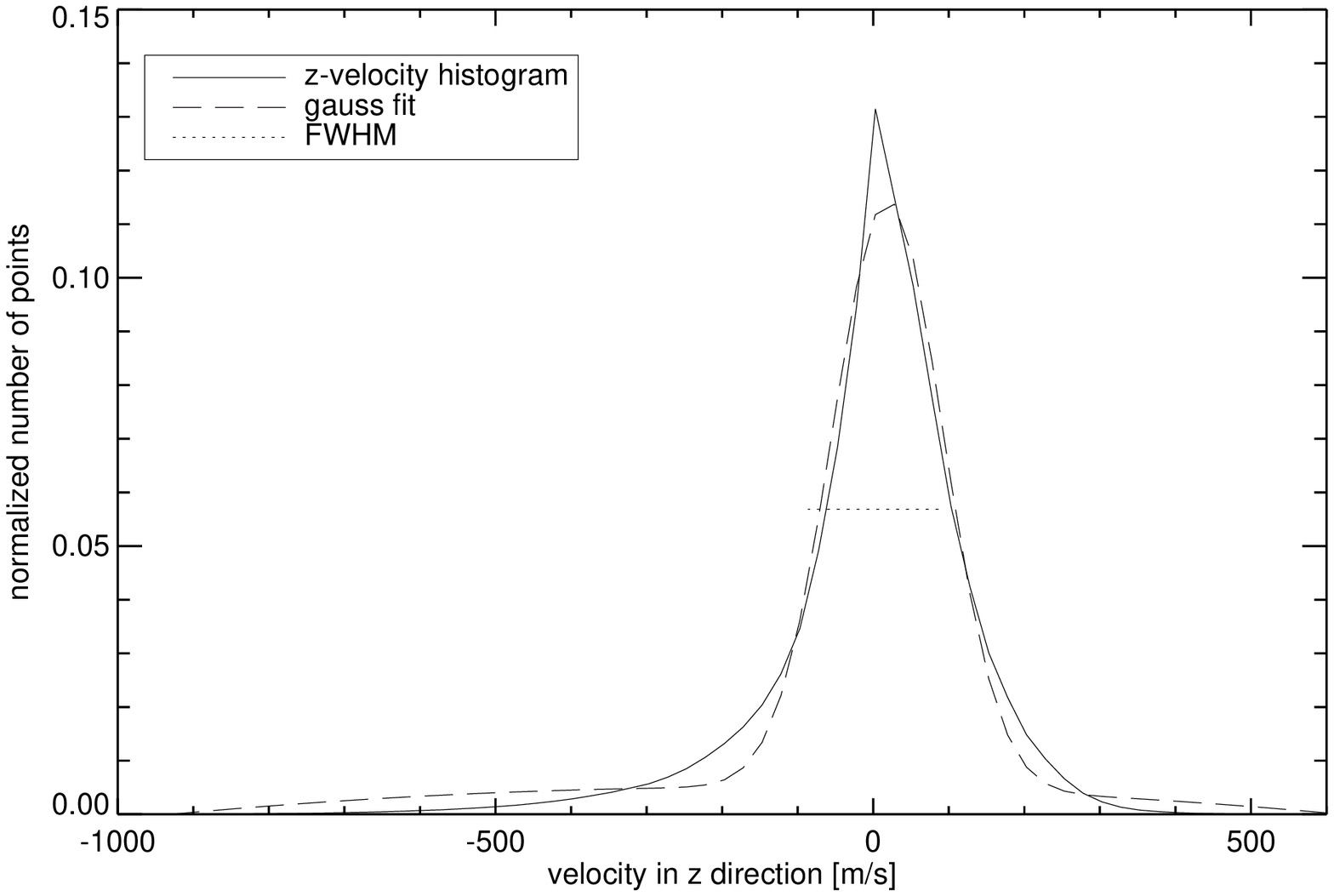}
 \caption{Histogram of the
 velocity distribution in vertical direction. The normalized number of points
 is plotted against the vertical velocity in m/s (solid line). 
 The Gaussian (dashed line) fits the velocity distribution and
 determined an FWHM value (dashed-dotted line), which is related to $\sigma$
 with $FWHM=2\sqrt{\ln{2}}\cdot\sigma$. The
 underlying model is located at $\teff=2800$\,K and $\log{g}=5$\,[cgs].}
\label{FWHMhisto}
\end{figure}
\begin{figure*}
\centering
 \includegraphics[width=18cm,bb=25 30 530 225]{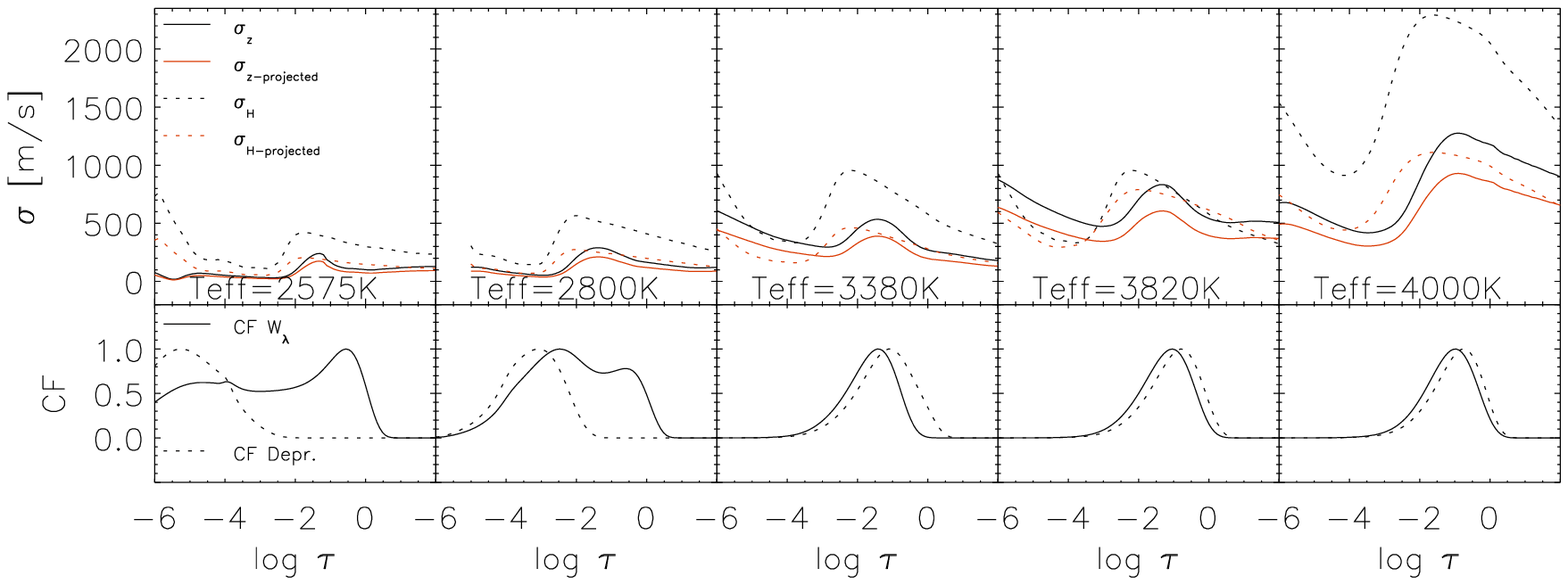}
 
\caption{Upper panel: Radial ($\sigma_z$) and horizontal ($\sigma_H$) 
components of the velocity dispersion are plotted
against the optical depth on a logarithmic scale.  Bottom panel: Each
bottom panel shows the contribution functions (CF) of an
\element[][]{FeH}-line at a wavelength of $9956.72$\,$\AA$. Equivalent width $W_{\lambda}$
(solid) and the depression at the line center (dashed)
of the line are plotted as a function of optical depth on a
logarithmic scale.  The models (from left to right) are located at
$\teff$ of $2800$\,K, $3380$\,K, $3820$\,K and $4000$\,K and a $\log{g}$
value of $5.0$, except the one with $\teff~=~3820$\,K
($\log{g}~=~4.9$), and the one with $\teff~=~4000$\,K ($\log{g}~=~4.5$)
[cgs].}
\label{FWHMteff}
\end{figure*}
\begin{figure*}
\centering
  \includegraphics[width=18cm,bb=25 30 530 225 ]{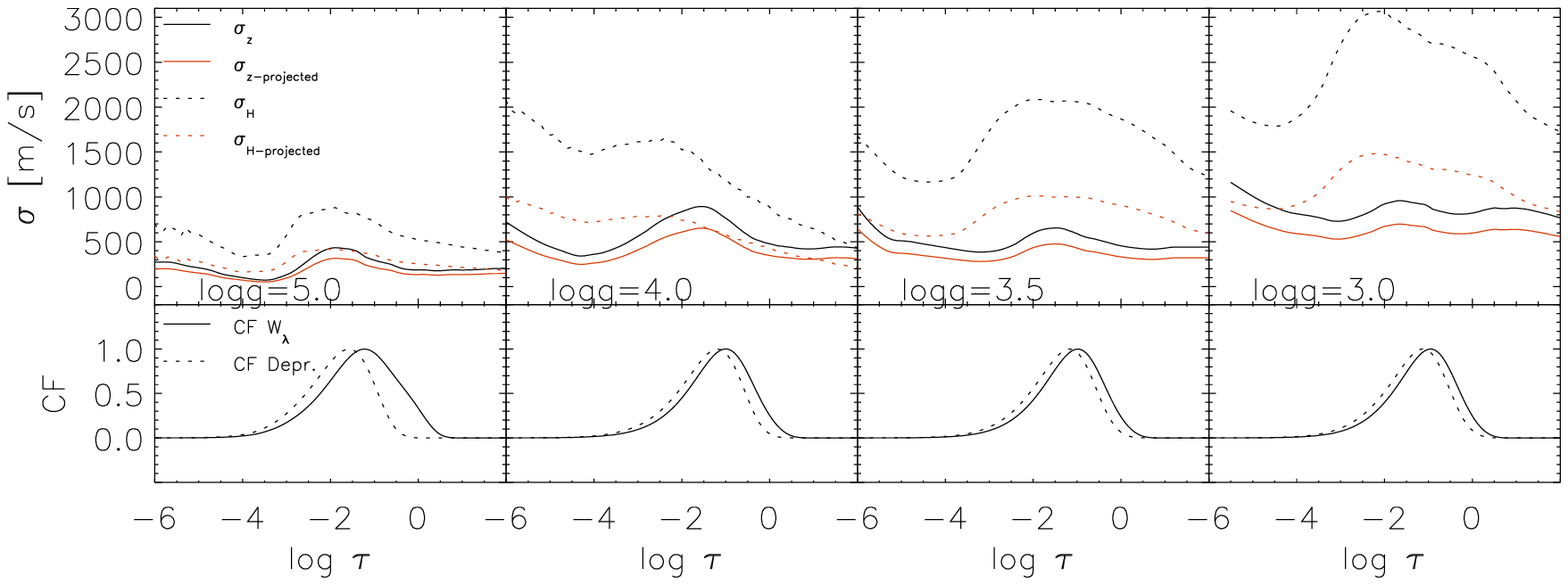}
 
\caption{Upper panel: Radial ($\sigma_z$) and horizontal ($\sigma_\mathrm{H}$) 
components of the velocity dispersion are plotted
against the optical depth on a logarithmic scale. Bottom panel: Each
bottom panel shows the contribution functions (CF) of an
\element[][]{FeH}-line at a wavelength of $9956.72$\,$\AA$. Equivalent width $W_{\lambda}$
(solid) and the depression at the line center (dashed)
of the line are plotted as a function of optical depth on a
logarithmic scale. The models are located at $\log{g}$ values from
left to right of 5.0 ($\teff~=~3275$\,K), 4.0 ($\teff~=~3315$\,K), 3.5
($\teff~=~3270$\,K) and 3.0 ($\teff~=~3240$\,K) [cgs].}
 \label{FWHMlogg}
\end{figure*}%
\begin{figure}
\centering
 \includegraphics[width=9cm,bb=15 55 580 660]{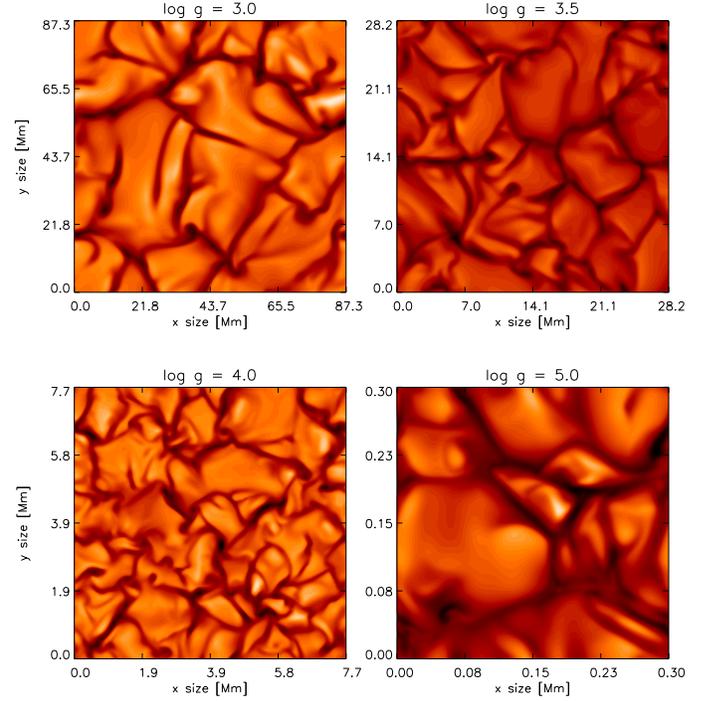}
 \caption{Horizontal cross-section around $\tau~\sim~1$ 
   of vertical velocity components. The models are located at $\log{g}$ values
   of $3.0$, $3.5$, $4.0$, and $5.0$\,[cgs] (from upper left corner to lower
   right, respectively).}
\label{loggcontour}
\end{figure}
\begin{figure}
\centering
 \includegraphics[width=9cm,bb=25 30 575 410]{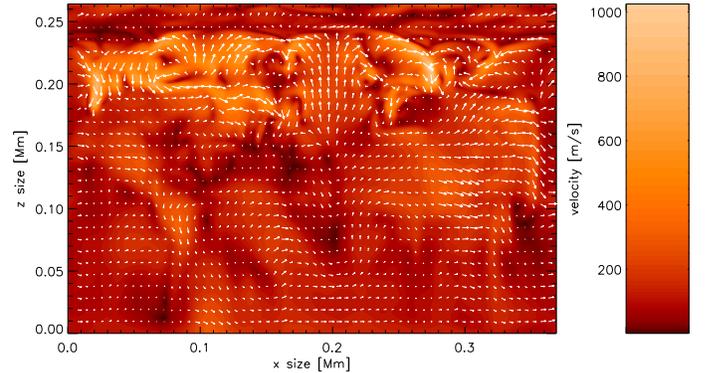}
 \caption{2D cross-section in x-z direction of the x-z velocity 
   in a model with $\log{g}~=~5.0$ [cgs] and
   $\teff~=~2800$\,K. Overplotted are x-z velocity vectors. The direction of
   the material flow is indicated by the velocity vectors. }
\label{T28zvelo}
\end{figure}
\begin{figure}
\centering
 \includegraphics[width=9cm]{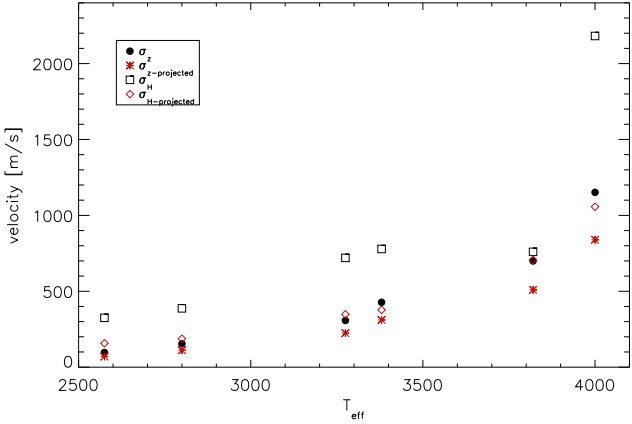}
 \caption{Plotted are the weighted projected and unprojected velocity dispersions
   of the horizontal and vertical component for models with different $\teff$. 
   The models are located at $\teff$ values of
   $2575$\,K, $2800$\,K, $3275$\,K, $3380$\,K, $3820$\,K and $4000$\,K and a
   $\log{g}$ value of 5.0, except the one with $\teff~=~3820$\,K 
   ($\log{g}~=~4.9$), and the one with $\teff~=~4000$\,K ($\log{g}~=~4.5$)
   [cgs].}
\label{teffabsvelos}
\end{figure}
\begin{figure}
\centering
  \includegraphics[width=9cm]{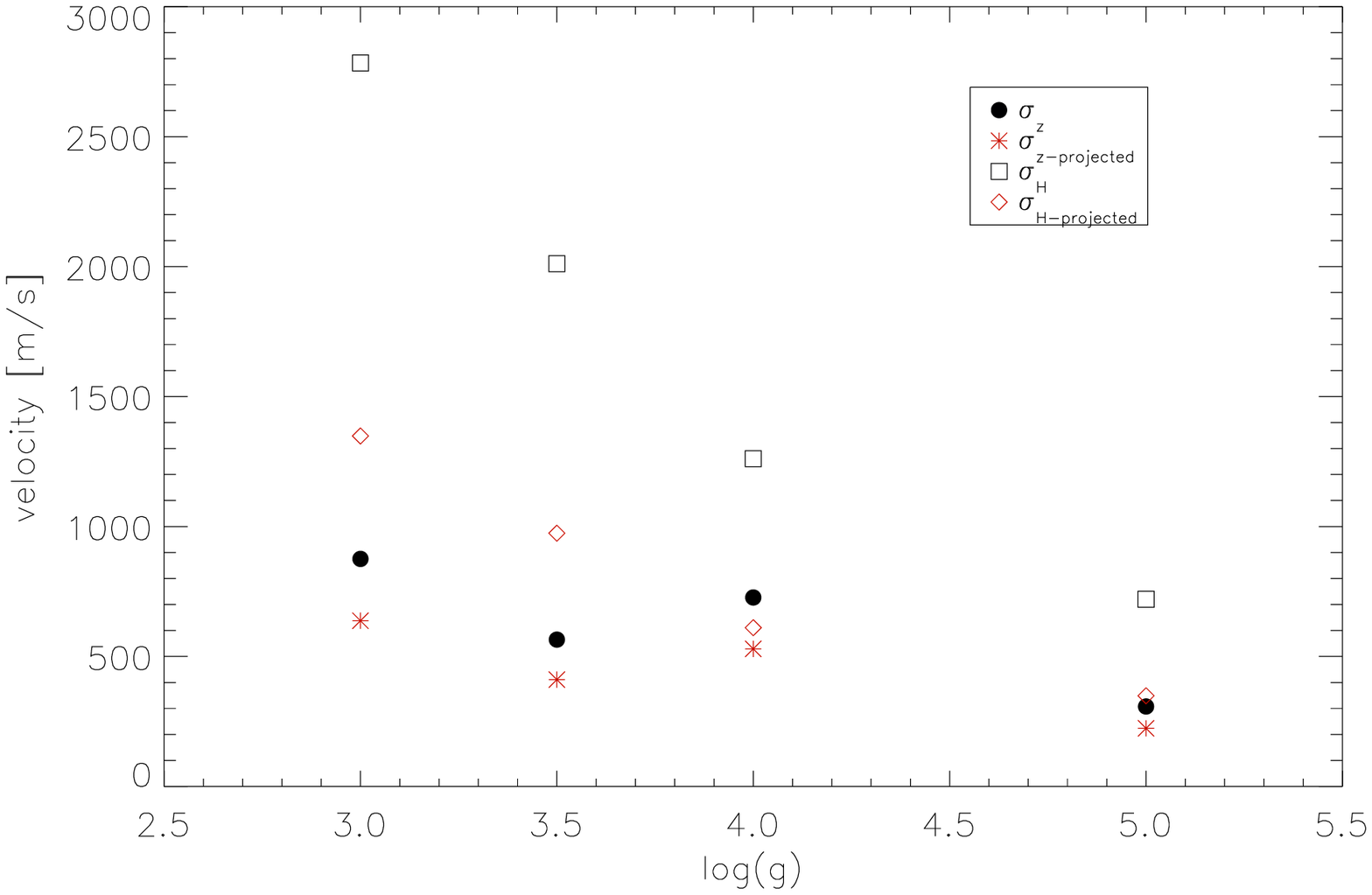}
 \caption{Plotted are the weighted projected and unprojected velocity dispersions
   of the horizontal and vertical component for
   models with different $\log{g}$. 
   The models are located at $\teff$ around
   $3300K$ and different $\log{g}$ values of 3.0 ($\teff~=~3240$\,K), 3.5
   ($\teff~=~3270$\,K), 4.0 ($\teff~=~3315$\,K) and 5.0 ($\teff~=~3275$\,K) [cgs].}
 \label{loggabsvelos}
\end{figure} 
We fit the histogram velocity distribution with a
Gaussian normal distribution function $G=\exp{(-(\frac{x}{\sigma})^2)}$ and
take the standard deviation $\sigma$ as a measure for the velocity dispersion
$\sigma$ in the models (see Fig.~\ref{FWHMhisto}) (The relation between the Gaussian
standard deviation $\sigma$ and the standard deviation $\sigma_{\mathrm{rms}}$
of the mean velocity is 
$\sigma=\sqrt{2} \cdot \sigma_{\mathrm{rms}}$). This is done for the
$\sigma_x$, $\sigma_y$, and $\sigma_z$ component of the velocity
vector for each horizontal plane from $\tau_{\mathrm{min}}$ to
$\tau_{\mathrm{max}}$, which are the highest and the deepest points,
in each model atmosphere. In
this way we obtained the height dependent velocity dispersion
$\sigma_{x,y,z}(\tau)$. We average over five model snapshots for a
better statistical significance.  In Figs.~\ref{FWHMteff} and
\ref{FWHMlogg}, the velocity dispersions for the horizontal
components $\sigma_\mathrm{H}=\sqrt{\sigma_x^2+\sigma_y^2}$ and vertical
component $\sigma_z$ are plotted against optical depth (black solid and dotted
lines). In the latter
figures, we identify the maxima around $\log{\tau}~=~-1$ of the
velocity dispersions as the region where the up-flowing motion spreads out in
horizontal direction and starts to fall back to deeper layers. We will call
this point, the ``convective turn-over point''. In a
horizontal 2D cross-section of the vertical velocities, somewhat below this
area, the up-flowing granulation patterns are
clearly visible (see Fig.\ref{loggcontour} for models with different
$\log{g}$ or \citet{2006A&A...459..599L}). In a vertical
2D cross-section one can see the behavior of the up-streaming material
(Fig.~\ref{T28zvelo} shows an example for a model with
$\teff=2800$\,K). The material starts moving upwards almost coherently
and before reaching higher layers (around $z~=~200$~km in
Fig.~\ref{T28zvelo}) of the atmosphere the vertical velocity dispersion
$\sigma_z$ becomes maximal. After that point, the material spreads out in
horizontal directions and starts falling down again. At this point,
the dispersion of horizontal velocities $\sigma_H$ becomes maximal (convective
turn-over point).

In Fig.\ref{FWHMlogg} we can see at lower
surface gravities that the maxima of the horizontal velocity dispersion are not centered
around a specified optical depth anymore; it spreads out in vertical direction
and span the widest range at $\log{g}~=~3.0$ [cgs]. The pressure stratification
changes, and the convective turn-over point moves to lower gas-pressure (not
shown here) but stays at almost constant optical depth between
$\log{\tau}~=~0$ and $\log{\tau}~=~-2$. With varying
temperature, the position of the convective
turn-over point stays at almost constant optical depth.
\subsection{Reduction of the 3D velocity fields} 
Commonly, micro- and macro-turbulence derived from spectroscopy
are interpreted as being associated with actual velocity fields present in the
stellar atmosphere. In our simulations no oscillations are induced
externally but small oscillations are generated in the simulations itself. The
velocity amplitudes of these oscillations reach maximal $10\,\%$ of the
convective velocities and have no significant influence on the
macro-turbulent velocity. We would also not expect global oscillations for
these objects, except for young stars with solar masses smaller than $0.1$\, $M_{\bigodot}$
induced by D-burning \citep{2005A&A...432L..57P}. In the following, we try to make the connection
between micro- and macro-turbulence and actual hydrodynamical velocity fields manifest
by considering the velocity dispersion determined directly from the
hydrodynamical model data, and compare it with the micro- and macro-turbulence
derived from synthesized spectral lines (see Sec.3.2). This connection is
algebraically not immediate, and we only apply a simple model for translating
the hydrodynamical velocities into turbulent velocities relevant to
spectroscopy (see appendix). When interpreting comparisons shown below
the very approximate nature of our model should be kept in mind.  In
this model, we include the geometrical projection of the components of
$\sigma_{x,y,z}$ to the line of sight of the observer. We also have to
consider the effect of limb-darkening of the stellar disk. For
each velocity dispersion component we calculate a projection factor which includes both,
geometrical projection and limb-darkening effects (more described in the
appendix). We take a limb-darkening coefficient of $0.4$ which
follows from the continuum from angle dependent line synthesis performed in
\texttt{LINFOR3D}. These simulations suggest that a linear limb-darkening law
with a limb-darkening coefficient of $0.4$ is suited to describe the
brightness variation. The projected velocity dispersions are also plotted in
Figs.~\ref{FWHMteff} and \ref{FWHMlogg} (red solid and dotted lines). The
reducing effect of this projection factor is stronger in the horizontal
components than in the vertical because the projected area at the limb of the
stellar disk, where $\sigma_H$ reaches its maximum value, is much smaller
than in the center where $\sigma_z$ has its maximum value. The influence of
limb-darkening is not strong and the dependence of the projection
factor from the limb-darkening coefficient is only small (described in more
detail in the appendix).
\subsubsection{Weighted velocities}
To investigate the influence
of broadening from the projected and unprojected velocity dispersion on spectral
lines, we use contribution functions for the equivalent width
$W_{\lambda}$ and the depression at the line center of an
\element[][]{FeH} line at $9956.7~\AA$
\citep{1986A&A...163..135M}. The line gains its $W_{\lambda}$ and
depression in the region between $\log{\tau}~=~1.0$ and $\log{\tau}~=~-4.0$,
i.e. that is the region of main continuum absorption caused by \element[][]{FeH} molecules. 
The maximum is roughly centered around $\log{\tau}~=~-1.0$ and moves
to slightly lower optical depth with lower temperatures (at
the lowest $\teff$ of $2575$\,K, the maximum is centered around $\log{\tau}~=~-2.0$) or
higher surface gravities.
The contribution function of $W_{\lambda}$
range over the region of the convection zone and reflects its 
influence on the line shape. Due to the latter fact, \element[][]{FeH} lines
are a good mean to explore the convective regions in M-dwarfs. In order to
measure the velocities in the region where the lines originate, we compute the
mean of the (projected) velocities weighted with the contribution function of $W_{\lambda}$.
\bq
\sigma_{\mathrm{weighted}}=\frac{\sum_{\tau=2}^{-6}\sigma_{\tau}\cdot
  CF_{\tau}}{\sum_{\tau=2}^{-6}CF_{\tau}}
\eq
The horizontal and vertical components of these weighted velocity dispersions
are plotted in Figs.~\ref{teffabsvelos} and \ref{loggabsvelos}. We can see an increase
of $\sigma_{H,z}$ with increasing effective temperature or decreasing surface gravity.
We can see again that the projected horizontal velocity dispersions are significantly smaller than
the unprojected ones due to the reasons mentioned above. The difference in the vertical component is
much smaller. The velocity dispersions for the vertical component range from a few hundred m/s for cool,
high gravity models to $\sim~1$\,km\,s$^{-1}$ for hot or low gravity models. The horizontal component range from
a $\sim~500$\,m\,s$^{-1}$ for cool, high gravity models to $\sim~2$\,km\,s$^{-1}$ for hot or low gravity models. 
We compare the total projected velocity dispersion $\sigma_{tot}=\sqrt{\sigma_H^2+\sigma_z^2}$ in 
Sec. 3.2 with micro- and macro-turbulent velocities in the classical sense.

The strong increase of the velocity dispersions in the atmospheres to higher layers
(Figs.~\ref{FWHMteff} and \ref{FWHMlogg}), which some
models show, is related to convective overshoot into formally stable
layers. These velocities are originated by waves excited by
stochastical fluid motions and by advective motions
\citep[][and references therein]{2002A&A...395...99L}. 
However, it will not affect the spectral lines, because
the lines are generated in the region between an optical depth of $\log{\tau}~=~1.0$
and $\log{\tau}~=~-4.0$. The lines on the model with $\teff=2575$\,K are an
exception, they are formed in the outermost layers of the model and it is not
possible to compute the full range of formation of these lines, because the
atmosphere is not extended enough. One has to keep this in mind when regarding
the line dependent results of this model later in this chapter.  

\subsection{Radial velocity shifts}
Due to the fact that in a convective motion the up-flowing area is larger,
because it is hotter and less dense, than the down-flowing one, one expect a
net shift of the velocity distribution to positive velocities. That means, the
net amount of up-flowing area with hotter temperature, i.e. more flux in
comparison to the down-flowing area, results in a blue shift in the rest
wavelength position of a spectral line \citep[see
e.g.][]{1982ARA&A..20...61D}.

To see how the area of up-flowing material affects the rest
wavelength position of a spectral line, we computed ten \element[][]{FeH}
spectral lines (more described in Sect.~4) on 3D models to measure the
displacement of the line positions. In order to determine the center of the
line, we used the weighted mean
$C=\frac{\sum F\cdot v}{\sum F}$ which account for the asymmetric line shape.
(To use the weighted mean is appropriate here, since we have no
noise in the computed data.) The line shifts of the flux and the
intensity are given in Tab.~\ref{tabshiftadv}. A negative value stands for a
blue shift, and a positive for a red shift.
The values for each model are the mean of five temporal snapshots. 
The absolute displacement of the flux and intensity in the $\log{g}$ series reflects the
dependence of the velocity fields on surface gravity, but for the $\teff$
series a connection is barely visible (see Figs.~\ref{teffabsvelos}
and \ref{loggabsvelos}). This could be due to the small geometrical size of
the atmospheres in the $\teff$ series (see Tab.~\ref{tab1}). Only the the one with $\teff=4000$\,K
shows a significant line shift and in this model the atmosphere is $1150$\,km
high due to the slightly smaller $\log{g}$ value of $4.5$\,[cgs]. At this point
we will not go in deeper analysis of this topic.

Since we only use
five snapshots, we are dealing with statistics of small numbers and hence a
big scatter in the results. This scatter $\sigma_{\mathrm{shift}}$ is in general one order 
lower than the shift of the line and the integrated jitter of the line, which is 
important for radial-velocity measurements, goes with $\sigma_{\mathrm{jitter}}=
\frac{\sigma_{\mathrm{shift}}}{\sqrt{N}}$, where N is the number of snapshots. Since in a star N is of 
the order of $10^6$, the jitter will be of the order of \,mm\,s$^{-1}$. 
   
We did not further investigate the effect of granulation patterns on the line
profiles but, as we will see below, the lines are almost Gaussian and show no
direct evidence for significant granulation effects.  
 
\begin{table*}
\caption{Displacement $\Delta v_{\rm{Flux}}$ and $\Delta v_{\rm{Intensity}}$
  in m/s of the position of an \element[][]{FeH} line from the rest wavelength
  on models with different $\teff$ (left side) and different $\log{g}$ (right
  side).} 
\centering
\begin{tabular}{rrrrrrrr}
\hline\hline
\multicolumn{1}{l}{$T_{eff}$} & \multicolumn{1}{l}{$\Delta v_{\rm{Flux}}$
  [m/s]} & \multicolumn{1}{l}{$\Delta v_{\rm{intensity}}$ [m/s]}&
\multicolumn{1}{l}{$v_{\rm{ad}}$  [m/s]} &
\multicolumn{1}{l}{$\log{g}$} & \multicolumn{1}{l}{$\Delta v_{\rm{Flux}}$
  [m/s]} & \multicolumn{1}{l}{$\Delta v_{\rm{intensity}}$ [m/s]}&
\multicolumn{1}{l}{$v_{\rm{ad}}$  [m/s]}  \\ 
\hline
      
2800 & $0$ &$2$    & $3750$ & 3.0 & $-56$ &$-25$& $6100$            \\ 
3275 & $-1$ &$3$   & $4000$ & 3.5 & $-47$ &$-36$& $5500$           \\ 
3380 & $-2$ &$2$   & $4850$ & 4.0 & $-16$ &$-7$ & $5400$           \\
3820 & $-1$ &$-5$  & $6000$ & 5.0 & $-1$ &$3$   & $4500$             \\
4000 & $-44$ &$-28$& $7000$               \\
\hline
\end{tabular}
\label{tabshiftadv}
\end{table*}

\subsection{Micro- and macro-turbulent velocities}
\label{mmvelos}
Due to the large amount of CPU-time required to compute 3D RHD models
and spectral lines on these models, we want study the necessity of 3D models
in the range of M-stars. Our goal is to compare the broadening effects of the 3D
velocity fields on the shape of spectral lines with the broadening in terms of
the classical micro- and macro-turbulence profiles \citep[see
  e.g.][]{1977ApJ...218..530G,2008oasp.book.....G}. The latter description is
commonly used in 1D atmosphere models like \texttt{ATLAS9} \citep{1970SAOSR.309.....K} or
\texttt{PHOENIX} and related line
formation codes. If the difference between 1D and
3D velocity broadening is small, the usage of fast 1D atmosphere codes to
simulate M-stars for comparisons with observations, e.g. to
determine rotational- or Zeeman-broadening, would be an advantage. 

If the size of a turbulent element is small compared to unit optical depth,
we are in the regime of the micro-turbulence. The micro-turbulent velocities 
might differ strongly from one position to another and have a
statistical nature. The broadening effect on spectral lines can be described
with a Gaussian which enters the line absorption coefficient
\citep{2008oasp.book.....G}. 
It can be treated similar to the thermal
Doppler broadening. The effect on the shapes of saturated lines is an enhancement of line wings
due to the fact that at higher velocities the absorption cross section increases and as a consequence
the equivalent width ($W_{\lambda}$) of the line is increased.
 
If the size of a turbulent element is large compared to unit optical depth
(maybe better of the same size as unit optical depth), we are in the regime of
the macro-turbulence. It can be treated similar to rotational broadening as a
global broadening of spectral lines. The effect is an increase of the line
width but the equivalent width remains constant.

As we saw before, the
velocity fields in the M-stars are not very
strong in comparison to the sound speed (see Tab.\ref{tabshiftadv}) and
one could expect that their influence on line shapes do not deviate strongly
from Gaussian broadening.

\begin{figure}[!h]
\centering
 \includegraphics[width=9cm,bb=70 30 530 400]{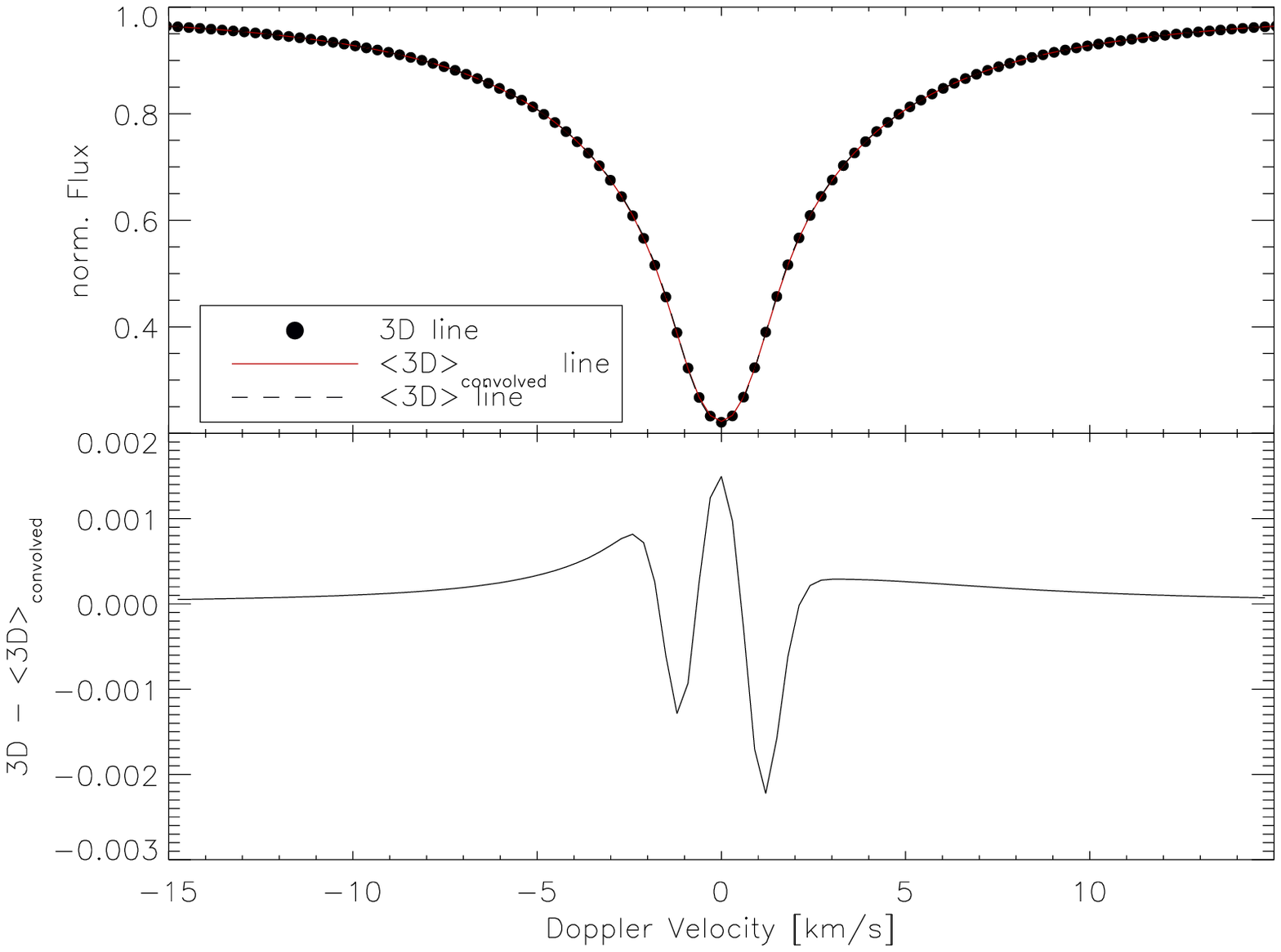}
 \includegraphics[width=9cm,bb=70 30 530 400]{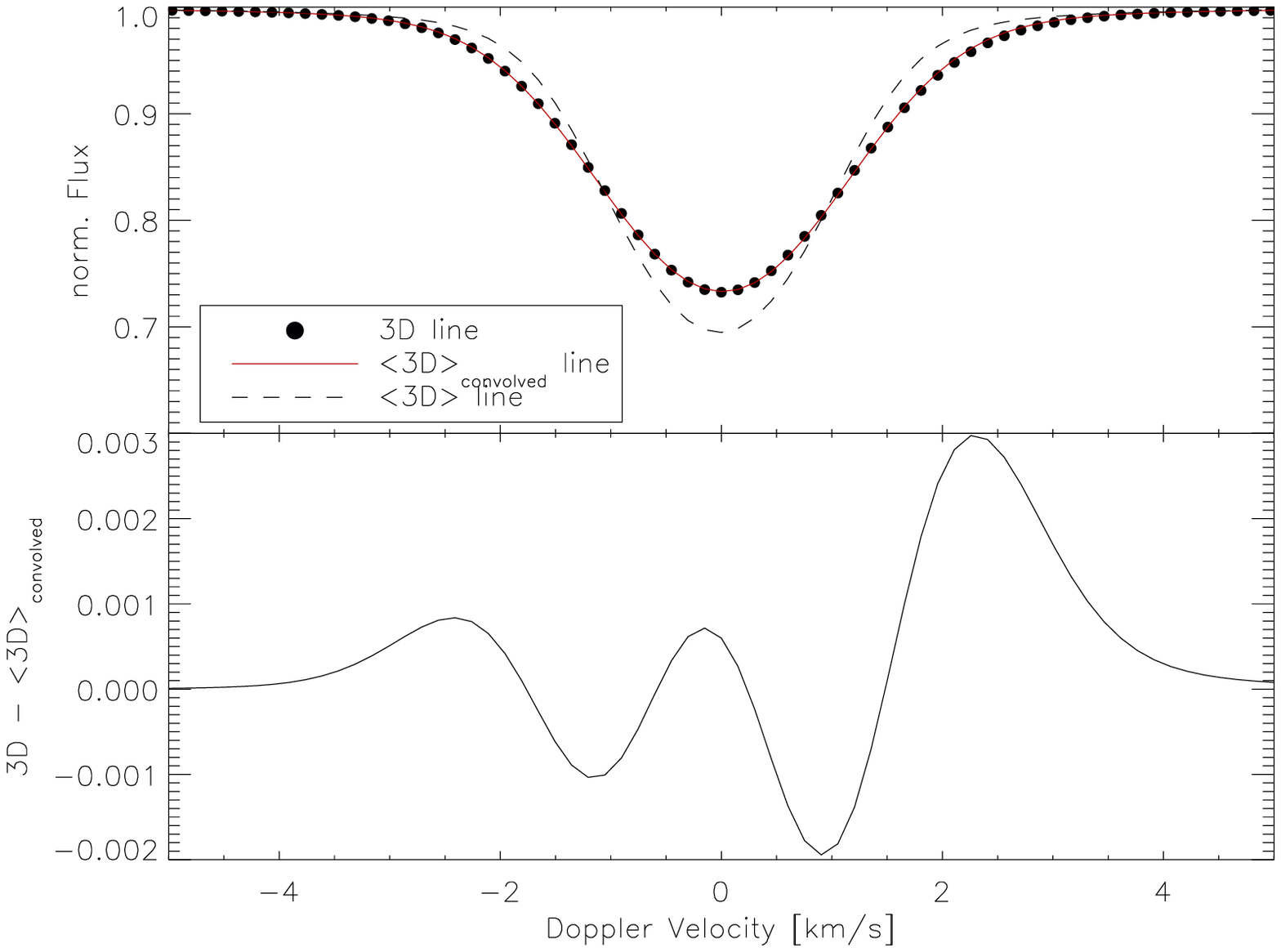}
 \caption{FeH lines for models with $\teff~=~2800$\,K, $\log{g}~=~5.0$ [cgs]
 (top) and $\teff~=~3820$\,K, $\log{g}~=~4.9$ [cgs] (bottom).  The upper panels
 show the 3D-line (dots) and the \mD$_{\mathrm{convolved}}$-line (solid line)
 which was broadened by a Gaussian profile. For comparison we plotted a
 \mD-line which was not broadened by any velocities (dashed line). In the
 lower panels the 3D-\mD$_{\mathrm{convolved}}$ residuals are plotted. One can
 see the asymmetry which stems from the line shifts due to convective
 motions. Note the different y-axis scale.}
\label{gausconv}
\end{figure}

We compared line broadening with the radial-tangential profile from
\citet{1975ApJ...202..148G} and a simple Gaussian profile and found that the
latter is a good approximation with an accuracy high enough for possible
determination of rotational- or Zeeman-broadening. Hence, in this
investigation we will assume Gaussian broadening profiles. That means, we
can assume a height independent isotropic velocity distribution for micro-
and macro-turbulent velocities. This is a
very convenient way to simulate the velocity fields. One would expect that
the anisotropic nature and the height dependence of the hydrodynamical
velocity fields have a significant influence on line shapes, so it is remarkable that
their influence on spectral lines can be described to a high
accuracy in this way. (At least in the investigated M-type stars.) In
Fig.~\ref{gausconv}, a few examples of \mD-\element[][]{FeH} spectral lines
were plotted, which were computed with a given micro-turbulent velocity
(determined below) and then convolved with a Gaussian broadening profile with a given
macro-turbulent velocity. The broadened \mD-\element[][]{FeH} lines fit the
3D-\element[][]{FeH} lines very well. The difference of the 1D and 3D centroid
($C=\frac{\sum F\cdot v}{\sum F}$) range in the order of m/s for small
velocity fields to $30-40$\,m\,s$^{-1}$ for strong velocity fields in hot M star models
or with low $\log{g}$. The error in flux is lower than $1\%$ (see
Fig.~\ref{gausconv}) this corresponds to an uncertainty in velocity, for example rotational
velocity, of less than $150$\,m\,s$^{-1}$ depending on the position in the line.
It is also visible in the latter figures that at low
effective temperature effects from velocity broadening are not visible in
comparison with an unbroadened \mD-line in which the v.d.Waals broadening is
dominant. At higher effective temperatures, the difference between broadened
and unbroadened \mD-lines is clearly visible. We found that in the range of
M-type stars, 1D spectral synthesis of
\element[][]{FeH}-lines using
micro- and macro-turbulent velocities in the classical description is
absolutely sufficient to include the effects of the velocity fields. 
In the following we will determine the velocities needed.
\subsubsection{Determination of micro- and macro-turbulent velocities}
The investigation of the micro-turbulent velocities was done with the
\emph{curve of growth} (CoG) method \citep[e.g.,][]{2008oasp.book.....G}. We
artificially increase the line strength of an absorption-line (increase the
$\log{gf}$ value) which in turn increase the saturation of the line and its
influence of the micro-turbulent velocity, which results in an enhancement of
$W_{\lambda}$.  In order to determine micro-turbulent-velocities, we use
\ion{Fe}{I}- and \element[][]{FeH}-lines produced on \mD-models with different
micro-turbulent velocities (there are no differences in micro-turbulent
velocities between both types of lines), i.e. for each $\log{gf}$-value we
compute a \mD-line with micro-turbulent velocities between $0$\,km\,s$^{-1}$
and $1$\,km\,s$^{-1}$ in $0.125$\,km\,s$^{-1}$ steps. In this way we obtained
CoGs for $9$ different micro-turbulent velocities. We compare the equivalent
widths in the CoGs with the ones which was computed on the 3D-models and
selected the velocity of the CoG which fits the 3D CoG best in the sense of
$\chi^2$-residuals.  
\begin{figure}[!h]
\centering
 \includegraphics[width=8cm,bb=80 20 540 386]{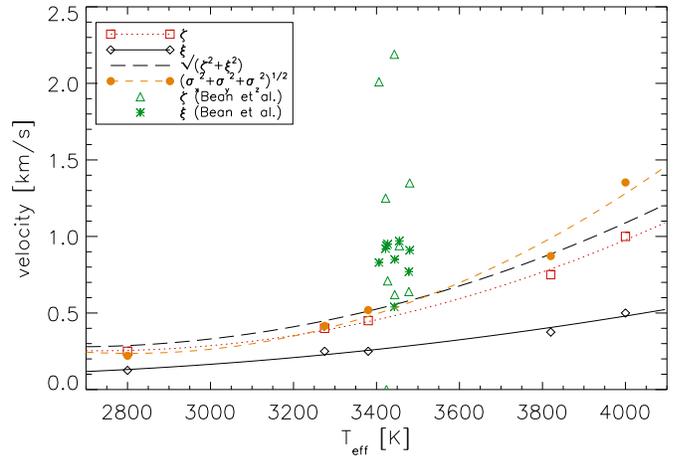}
 \caption{Macro- (solid) and micro- (dotted) turbulent velocities and the sum
   of both (long dashed line) as a function of $\teff$.  The data points are
   fitted by a second order polynomial and the sum of micro- and
   macro-turbulent velocities is given by the sum of the fits. For comparison,
   velocities from Bean et al. and the total projected weighted 3D velocity
   dispersions are plotted, too (short dashed line). The models are located at
   $\teff$ values of $2800$\,K, $3275$\,K, $3380$\,K, $3820$\,K and $4000$\,K and a
   $\log{g}$ value of 5.0, except the one with $\teff~=~3820$\,K
   ($\log{g}~=~4.9$), and the one with $\teff~=~4000$\,K ($\log{g}~=~4.5$)
   [cgs].}
\label{teffvelos}
\end{figure}
\begin{figure}[!h]
\centering
  \includegraphics[width=8cm,bb=80 20 540 386]{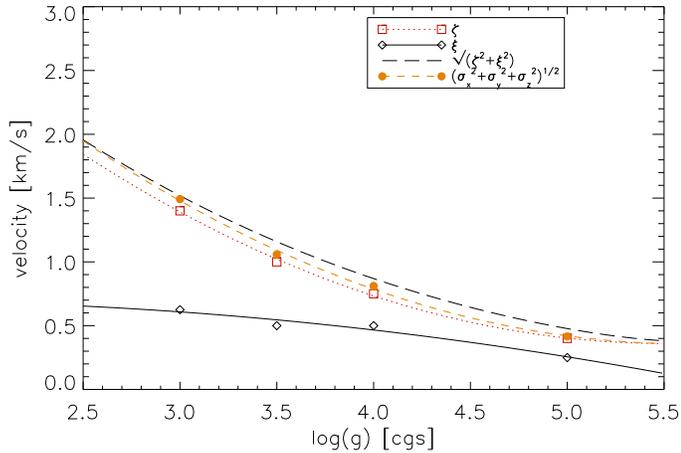}
 \caption{Macro- (solid) and micro- (dotted) turbulent velocities and the sum
   of both (long dashed line) as a function of $\log{g}$. The data points are
   fitted by a second order polynomial and the sum of micro- and
   macro-turbulent velocities is given by the sum of the fits. For comparison,
   the total projected weighted 3D velocity dispersions is plotted, too (short
   dashed line).  The models are located at $\teff$ around $3300$\,K and
   different $\log{g}$ values of 3.0 ($\teff~=~3240$\,K), 3.5
   ($\teff~=~3270$\,K), 4.0 ($\teff~=~3315$\,K) and 5.0 ($\teff~=~3275$\,K)
   [cgs].}
 \label{loggvelos}
\end{figure}
The macro-turbulent velocities are determined by computing a \mD-model
\element[][]{FeH} line (at $\lambda=9956.7~\AA$)
including the micro-turbulent velocities.
This line is then broadened with a Gaussian profile and different
velocities until it matches the 3D-model line.
 
The dependence of the micro- ($\xi$) and macro- ($\zeta$) turbulent velocities on surface gravity and
effective temperature is plotted in Figs.~\ref{teffvelos} and
\ref{loggvelos}. The macro-turbulent velocities in both cases show a quadratic
dependence, and we can fit them with a second order polynomial. The
micro-turbulent velocities could be fitted by a linear function or a second
order polynomial. We decided to
use the second order polynomial, too.

Micro- and
macro-turbulence velocities both show a similar dependence on surface gravity and
effective temperature, which implies that there is a direct connection between both.
A comparison of the macro ($\zeta$)- and micro ($\xi$)-turbulent velocities
with the sum of both ($\sqrt{\zeta^2+\xi^2}$)
and the total projected weighted velocity dispersion $\sigma_{\mathrm{tot}}$ (see Sec. 3) is also shown in
Figs.~\ref{teffvelos} and \ref{loggvelos}. 
The total projected weighted velocity dispersion (see Sec.3) is very similar
to the macro-turbulent velocities 
and in most cases smaller than the sum of micro- and macro-turbulent velocity. It is remarkably,
that it is possible, with this simple description of the total projected weighted velocity, 
to describe the broadening influence of the hydrodynamical velocity fields on spectral lines
in comparison with the classical micro- and macro-turbulent description.

In order to obtain a good estimate of the line profile in 1D spectral
line synthesis, knowledge of the micro- and macro-turbulent
velocities is very important. Otherwise one could underestimate the equivalent
width or the line width and hence obtain a wrong line depth.

We compare our micro- and macro-turbulent velocities to observational results
from
\citep[see][]{2006ApJ...653L..65B,2006ApJ...652.1604B,2007PhDT........18B}.
Our value of the macro-turbulent velocities are roughly of the same order. The
higher macro-turbulent velocities from Bean et al. possibly contain rotational
broadening, but the Bean et al. micro-turbulent velocities are roughly a
factor two or three higher than ours. These velocities are obtained from
observed spectra by the authors of the afore-mentioned papers using spectral
fitting procedures. In detail, they used \texttt{PHOENIX} atmosphere models
and the stellar analysis code \texttt{MOOG} \citep{1973PhDT.......180S}. One
has to keep in mind that the empirical determination of micro-turbulence is
also endangered to suffer from systematic errors. For most of the lines
that Bean and collaborators employ (line data from \citet{2000A&AS..142..467B}),
the v.d.Waals damping constant is available. However, in some cases not, then
usually Uns\"old's hydrogenic approximation is applied to calculate the value,
and different authors use significantly different enhancement factors changing
its value. This illustrates the level of uncertainty inherent to this
approach. For instance, \citet{1996MNRAS.283..821S} use an enhance factor of
5.3 (for the resulting $\gamma_6$ values) for their \ion{Fe}{I} lines, while
Bean et al. prefer 2.5 for \ion{Ti}{I} lines \citep{2007PhDT........18B}. 
To investigate the detailed influence of
v.d.Waals broadening on determination of micro-turbulence velocities is very
interesting, but goes beyond the scope of this paper. We want to point out, that
it is perceivable that uncertainties in the damping constant can introduce
significant systematic biases in the resulting value of spectroscopically
micro- and macro-turbulence and they could easily be overestimated. 

As mentioned above and illustrated 
Fig.~\ref{teffvelos}, our prediction of the micro-turbulence
grossly underestimates the micro-turbulence values measured by Bean et
al.. This might hint at deficits in the hydrodynamical modeling and we cannot
exclude the possibility that a process is missing in our 3D models leading to
a substantially higher micro-turbulence. But due to the argumentation above
and a comparison with the solar micro- and macro-turbulence, we argue
that the Bean et al. values for the micro-turbulence are too high. However,
before being able to draw definite conclusions the observational basis has to
be enlarged.

\section{$\teff$- and $\log{g}$-dependence  of  \element[][]{FeH} molecular lines}

In this section we study the dependence of \element[][]{FeH} molecular
lines on $\teff$ and $\log{g}$ in our 3D- and \mD-models. Again, the intention
is to identify multi-D effects which might hamper the use of the
\element[][]{FeH} line diagnostics in standard 1D analyses. For this purpose
we compute the \mD-lines with no micro- and macro-turbulence velocity. With
this method we can study the \element[][]{FeH} lines without any velocity
effects and can through direct comparison between 3D- and \mD-lines clearly
identify velocity induced effects. 

\subsection{\element[][]{FeH} line data}
\citet{1969PASP...81..527W} were the first to detect a broad molecular
absorption band around $991$\,nm in late M dwarfs. This band was later found in
S stars \citep{1972saim.conf..123W} and in sun spots.
\citet{1977A&A....56....1N} identified the Wing-Ford band as the $0-0$ band of
a $\element[][]{FeH}$ electronic transition. The $\element[][]{FeH}$ molecule
is well suited for the measurements mentioned in the introduction because of
its intrinsically narrow and well isolated spectral lines. These lines are
also an ideal tracer of line broadening in M-stars due to convection or very
slow rotation \citep{2007A&A...467..259R}. Since \element[][]{FeH} lines were
not very commonly used for the interpretation of stellar spectra in the past,
only little data is recorded in the literature. With the work of
\citet{2003ApJ...594..651D}, it is possible to determine the $\log{gf}$ value
and the transition energies. For the partition function, a combination of the
tabulated function in \citet{2003ApJ...594..651D} and an analytically
determined one from the Eq.$1$ in \citet{1984ApJS...56..193S} is used.

Due to the high atmospheric pressures, v.d.Waals broadening is often
significant in cool M-type dwarfs. No detailed calculations exist for the v.d.Waals
broadening of \element[][]{FeH} molecular lines. Lacking a more accurate
treatment, we follow the approximate approach of \citet{1996MNRAS.283..821S}
and apply Uns\"old's hydrogenic approximation, albeit -- different from
Schweitzer and co-workers -- we do not apply an enhancement factor to the
calculated broadening constant $C_6$.

The ionization energy of the \element[][]{FeH} molecule enters the calculation
of $C_6$ which, unfortunately, is not known. Only the dissociation energy
is published of $1.59\pm0.08$\,eV at 0\,K \citep[e.g.][]{2003ApJ...594..651D}.
To derive an estimate of the ionization energy, we heuristically compare the
ionization and dissociation energies of a large number of hydrides taking data
from \citet{1963ApJ...138..778W}.  We find an approximately linear
relationship between the ionization and the dissociation energies of hydrides.
For \element[][]{FeH} we obtain an ionization energy of $6$\,eV for the known
dissociation energy from a linear fit.  Computation of this value from the
ionization-potential of Fe and the dissociation energies of \element[][]{FeH}
and \element[][]{FeH}$^+$ yields $7.3$\,eV (Bernath 2008, private
communication), which is compatible with our value considering our rather
crude procedure.  The difference in FWHM of synthesized
\element[][]{FeH}-lines in the range between 6 and $7.3$\,eV amounts to
$\approx 25$\,m/s. This uncertainty is quite acceptable in comparison to the
total broadening of typically several 100\,m\,s$^{-1}$ in this investigation.

\subsection{An ensemble of 3D- and \mD-\element[][]{FeH} lines}

\begin{figure}
\centering
 \includegraphics[width=9cm,bb= 15 30 580 410 ]{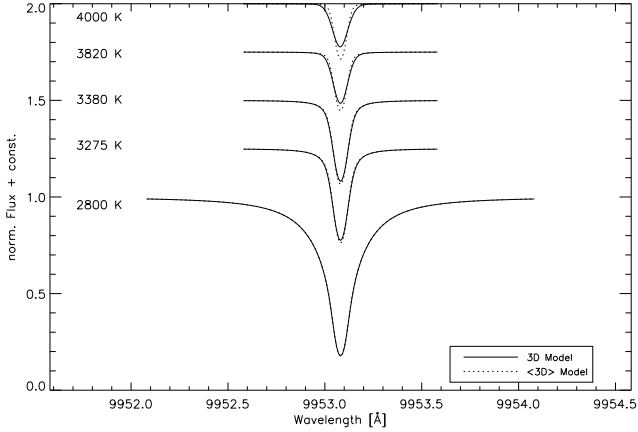}
 \caption{\element[][]{FeH} lines with constant $\log{g}$ and varying $\teff$.
  Plotted are pairs of 3D- and \mD-lines shifted by a constant for better
  visibility. The models are located at $\teff$ values of $2800$\,K, $3275$\,K,
  $3380$\,K, $3820$\,K and $4000$\,K and a $\log{g}$ value of 5.0, except the one
  with $\teff~=~3820$\,K ($\log{g}~=~4.9$), and the one with $\teff~=~4000$\,K
  ($\log{g}~=~4.5$) [cgs].}
\label{tefflines}
\end{figure}

\begin{figure}
\centering
  \includegraphics[width=9cm,bb= 15 30 580 410]{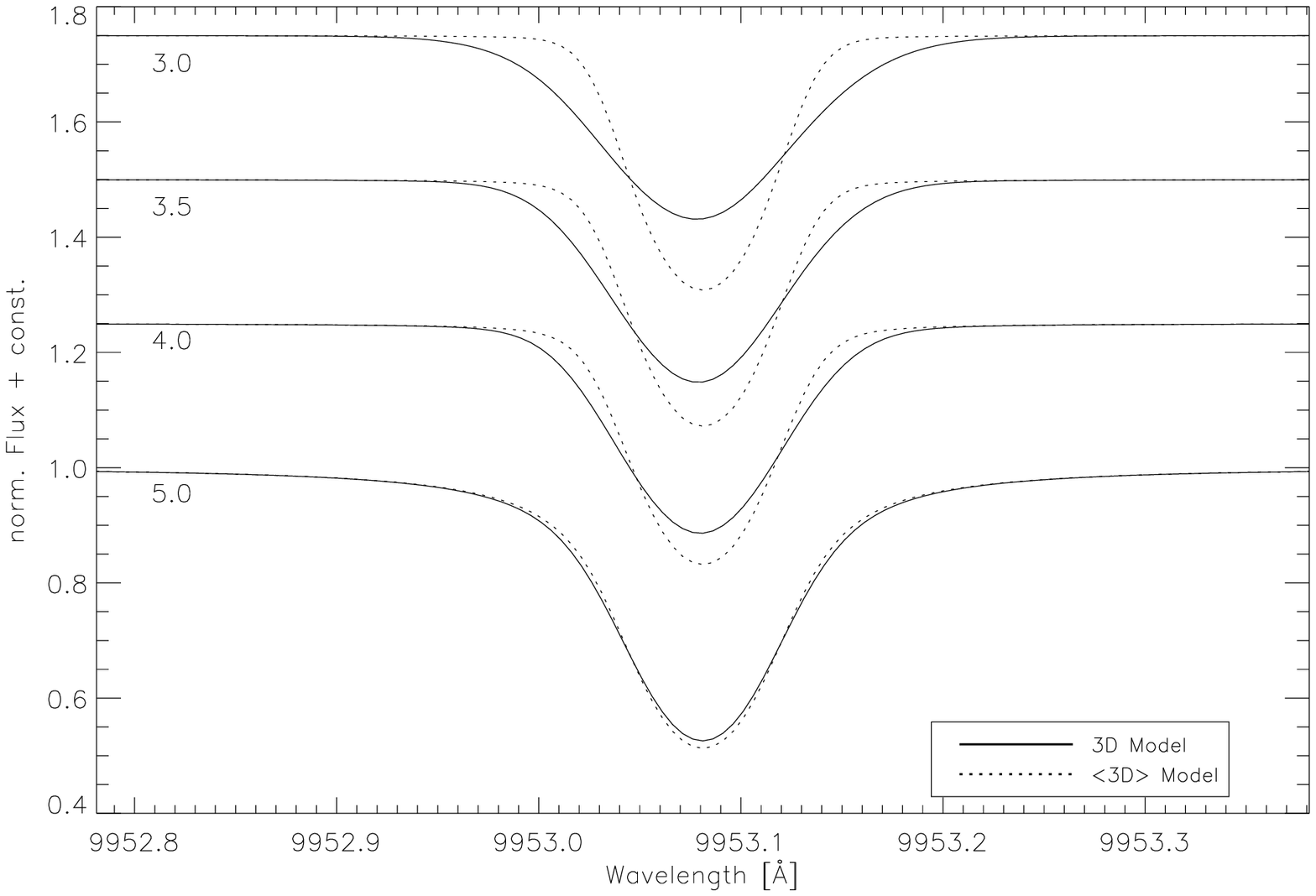}
 \caption{\element[][]{FeH} lines wit constant $\teff$ and varying $\log{g}$. 
  Plotted are pairs of 3D- and \mD-lines shifted by a constant for better
  visibility.
  The models are located at a $\teff$ around
  $3300K$ and different $\log{g}$ values of 3.0 ($\teff~=~3240$\,K), 3.5
  ($\teff~=~3270$\,K), 4.0 ($\teff~=~3315$\,K) and 5.0 ($\teff~=~3275$\,K)
  [cgs]. The large differences between 3D- and \mD-lines stem from the
  hydrodynamical velocity fields in the 3D models.}
 \label{logglines}
\end{figure}

We investigate ten \element[][]{FeH} lines between $9950$\,\AA\ and
$9990$\,\AA\ chosen from \cite{2006ApJ...644..497R} (see Tab.~\ref{tab2}). We
choose lines from different branches (Br), orbital angular momentum $\omega$, and rotational
quantum number $J$. The wavelengths in Tab.~\ref{tab2} are given in vacuum
and $E_l$ is the lower transition energy. While not directly relevant in the present context,
because we do not study the effects of magnetic fields, we note that five lines
are magnetically sensitive and five insensitive. We performed the line
synthesis for fixed abundances on the \texttt{CO$^5$BOLD} atmosphere models listed in
Tab.~\ref{tab1}. The spectral resolution is $ R ~ \approx ~
2\cdot 10^6 $ ($\equiv 5\cdot 10^{-3} $ \AA) corresponding to a Doppler
velocity of $v~\approx~150$\,m\,s$^{-1}$ at the wavelength of the considered lines
($\sim~9950~ \AA$).
Figures~\ref{tefflines} and
\ref{logglines} illustrate the strong influence of surface gravity and effective
temperature on the line shape for the 3D-models. In Fig.~\ref{tefflines}, one
can see that for both, \mD- and 3D-lines, the line depth, line width
and equivalent width $W_{\lambda}$ decrease strongly with increasing effective
temperature. The decrease of $W_{\lambda}$ is due to
stronger dissociation of the \element[][]{FeH} molecules at higher
temperatures, i.e the number of \element[][]{FeH} molecule absorber
decreases. Differences in the line shape between 3D- and \mD-lines with
changing temperature are hardly visible. To higher $\teff$ values, the
influence of broadening on the 3D lines due to velocity fields is slightly
visible and not covered by thermal and v.d.Waals broadening anymore. To cooler
temperatures, the velocity fields decrease and the differences between the
\mD- and 3D-line shapes vanishes. The v.d.Waals broadening is larger
then the thermal broadening or that from the small velocity fields in the RHD models. In the
model with $\teff~=~2800$\,K, the lines start to become saturated. The
\element[][]{FeH} lines in the z-band at effective temperatures below
$\sim~2600$\,K become
too saturated and too broad for investigations of quantities like magnetic
field strength or rotational broadening below $10$\,km\,s$^{-1}$.

The differences between 3D- and \mD-line shapes of the $\log{g}$-series in
Fig.~\ref{logglines} is more obvious than in the $\teff$ case. The differences
in line depth and line width become significant
with decreasing $\log{g}$. The lines in the 3D-models are significantly
broadened due to the velocity fields in the RHD models, hence the line width is
larger and the line depth lower. As we saw in Chapter~3, these velocity fields
increase with decreasing $\log{g}$ and could be described in the 1D case in
terms of macro- and micro-turbulent velocities.

The \mD-lines become slightly shallower and narrower towards smaller $\log{g}$.
The equivalent width of the lines decreases with decreasing $\log{g}$ due to
decreasing pressure and hence decreasing concentration of of \element{FeH} molecules.
Also the v.d.Waals broadening looses its influence to lower pressures and the
line width decreases.
We point out, that we use in all models the same solar like chemical
compositions, hence the concentration of \element{Fe} and \element{H}
stays the same. The creation of \element[][]{FeH} also depends on the number
of H$_2$-molecules, which become larger towards lower temperatures and
will be important in cool models.

The slightly different effective
temperatures in the models with different $\log{g}$ (see Tab.~\ref{tab1})
affect the line depths as well. If the effective temperatures would be the
same in the $\log{g}$-models, one could expect a monotonic behavior in
decreasing line strength for decreasing surface gravity in the
\mD-lines. However, because the model with $\log{g}~=~3.0$ is
cooler, the line depth is deeper than that of the one with
$\log{g}~=~3.5$.  In the following analysis we will correct the FWHM,
$W_{\lambda}$, and the line depth of the lines on models with different
$\log{g}$ for their slightly different effective temperatures.

The ten \element[][]{FeH} lines behave all in the same way as the presented ones. We
do not see any effect of different excitation potentials or $\log{gf}$~values 
in the line shapes that cannot be explained by their different height of formation. Thus,
we expect that we can exclude an extraordinary interaction between these quantities and
effective temperature or surface gravity. We will quantify this preliminary
result in the next section.
\begin{table}[!h]
\centering
\caption{Several quantities of the investigated \element[][]{FeH} lines
 \citep{2006ApJ...644..497R}.}
\begin{tabular}{lrrrrrr}
\hline\hline
 $\lambda_{rest}$ [\AA] & $\log{gf}$ & $E_l$~[eV] & $Br.$ & $J$ & $\Omega$ & $magn.~sen.$ \\ 
\hline
9953.08 & -0.809 & 0.156 & R & 10.5 & 1.5 & weak \\ 
9954.00 & -2.046 & 0.199 & P & 16.5 & 2.5-3.5 & strong \\ 
9956.72 & -0.484 & 0.375 & R & 22.5 & 3.5 & strong \\ 
9957.32 & -0.731 & 0.194 & R & 12.5 & 1.5 & weak \\ 
9973.80 & -0.730 & 0.196 & R & 12.5 & 1.5 & weak \\ 
9974.46 & -1.164 & 0.108 & R & 4.5 & 0.5 & weak \\ 
9978.22 & -1.190 & 0.030 & Q & 2.5 & 2.5 & strong \\ 
9979.14 & -1.411 & 0.093 & R & 2.5 & 0.5 & strong \\ 
9981.46 & -1.006 & 0.130 & R & 6.5 & 0.5 & weak \\ 
9982.60 & -1.322 & 0.035 & Q & 3.5 & 2.5 & strong \\ 
\hline
\end{tabular}

\label{tab2}
\end{table} 


\subsection{\element[][]{FeH} Line shapes}
\label{sec:lineshapes}
\begin{figure}
\centering
 \includegraphics[width=9cm,bb=65 255 550 630]{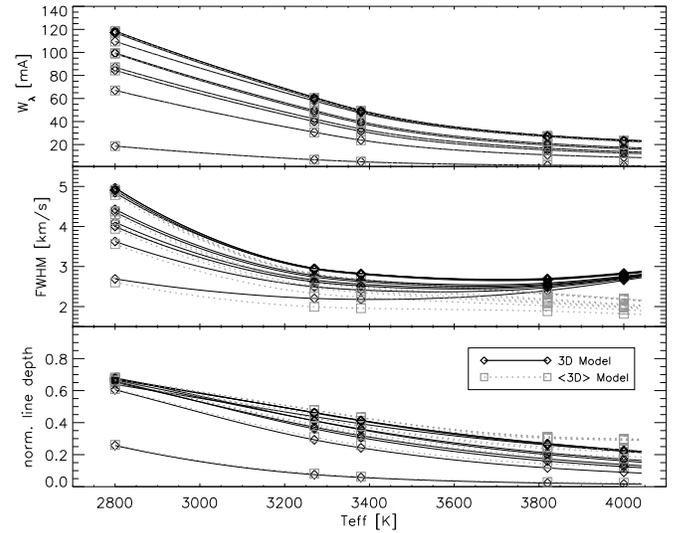}
 \caption{Run of $W_{\lambda}$ (top), FWHM (middle) and the line depth
  (bottom) of ten \element[][]{FeH}-lines (see Tab.~\ref{tab2}) on models with
  different $\teff$. Squares are the data points for the \mD-models and diamonds for
  the 3D-models. $W_{\lambda}$ and FWHM are on a logarithmic ordinate for better
  visibility. We connect the data points of
  each line with fitting functions (see text) to guide the eye. Plotted are the quantities of
  the 3D-models (black solid lines) and the \mD-models (gray dotted
  lines). The models are located at $\teff$ values of $2800$\,K, $3275$\,K,
  $3380$\,K, $3820$\,K and $4000$\,K and a $\log{g}$ value of 5.0, except the one
  with $\teff~=~3820$\,K ($\log{g}~=~4.9$), and the one with $\teff~=~4000$\,K
  ($\log{g}~=~4.5$) [cgs].}
\label{teffdata}
\end{figure}

\begin{figure}
\centering
  \includegraphics[width=9cm,bb=65 255 550 630]{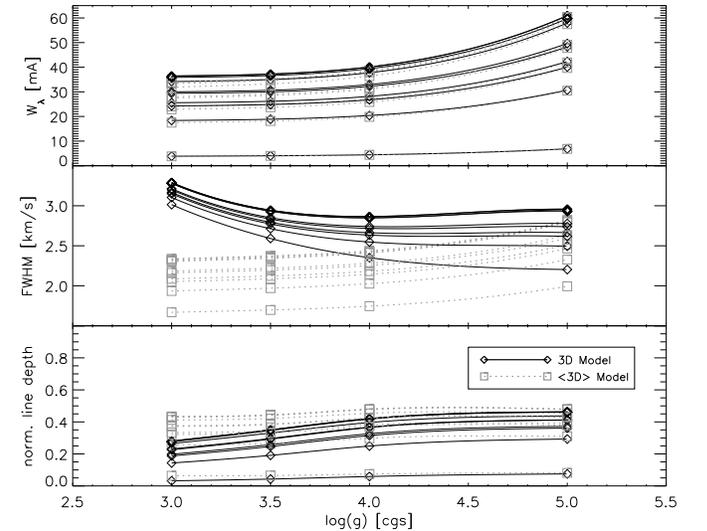}
 \caption{Run of $W_{\lambda}$ (top), FWHM (middle) and line depth (bottom) of
  ten \element[][]{FeH}-lines (see Tab.~\ref{tab2}) on models with different
  $\log{g}$. Squares are the data points for the \mD-models and diamonds for
  the 3D-models. We
  connect the data points of each line with
  fitting functions (see text) to
  guide the eye. Plotted are the quantities of the 3D-models
  (black solid lines) and the \mD-models (gray dotted lines). All quantities
  are corrected to a $\teff$ of $3275$\,K.}
 \label{loggdata}
\end{figure}

To quantify the visual results of Fig.~\ref{tefflines} and
Fig.~\ref{logglines}, we measured $W_{\lambda}$, the FWHM, and
the line depth of the ten investigated \element[][]{FeH}-lines (see
Tab.~\ref{tab2}). We compare 3D- and \mD-lines with each other to study
the effects of the velocity fields in the 3D models and to explore the
behavior of the \element[][]{FeH} without broadening effects from the
hydrodynamical motions. A run of these quantities is plotted in Fig.~\ref{teffdata}
and Fig.~\ref{loggdata}.

As we mentioned above, we have to correct the line
quantities from models with changing $\log{g}$ for their slightly different
$\teff$.  In Figs.~\ref{tefflines} and \ref{teffdata} one can see how the
investigated quantities depend on $\teff$. We determined spline fits $\wp$ for 
the three quantities of each line. For these fitting functions $\wp$ we took into
account all five
different effective temperatures. In order to correct
the line quantities to a reference temperature of $\teff^{\mathrm{Ref.}}=3275$\,K, we
use a correction factor
$\gamma=\frac{\wp_{\mathrm{quant.}}(\teff^{\mathrm{Ref.}})}{\wp_{\mathrm{quant.}}(\teff.)}$ and multiply
the quantity for the $\log{g}$-model with $\gamma$. This gives us the value of
the quantity for a $\log{g}$-model which would have $\teff=3275$\,K.

\begin{figure}
\centering
  \includegraphics[width=9cm,bb= 15 30 580 410]{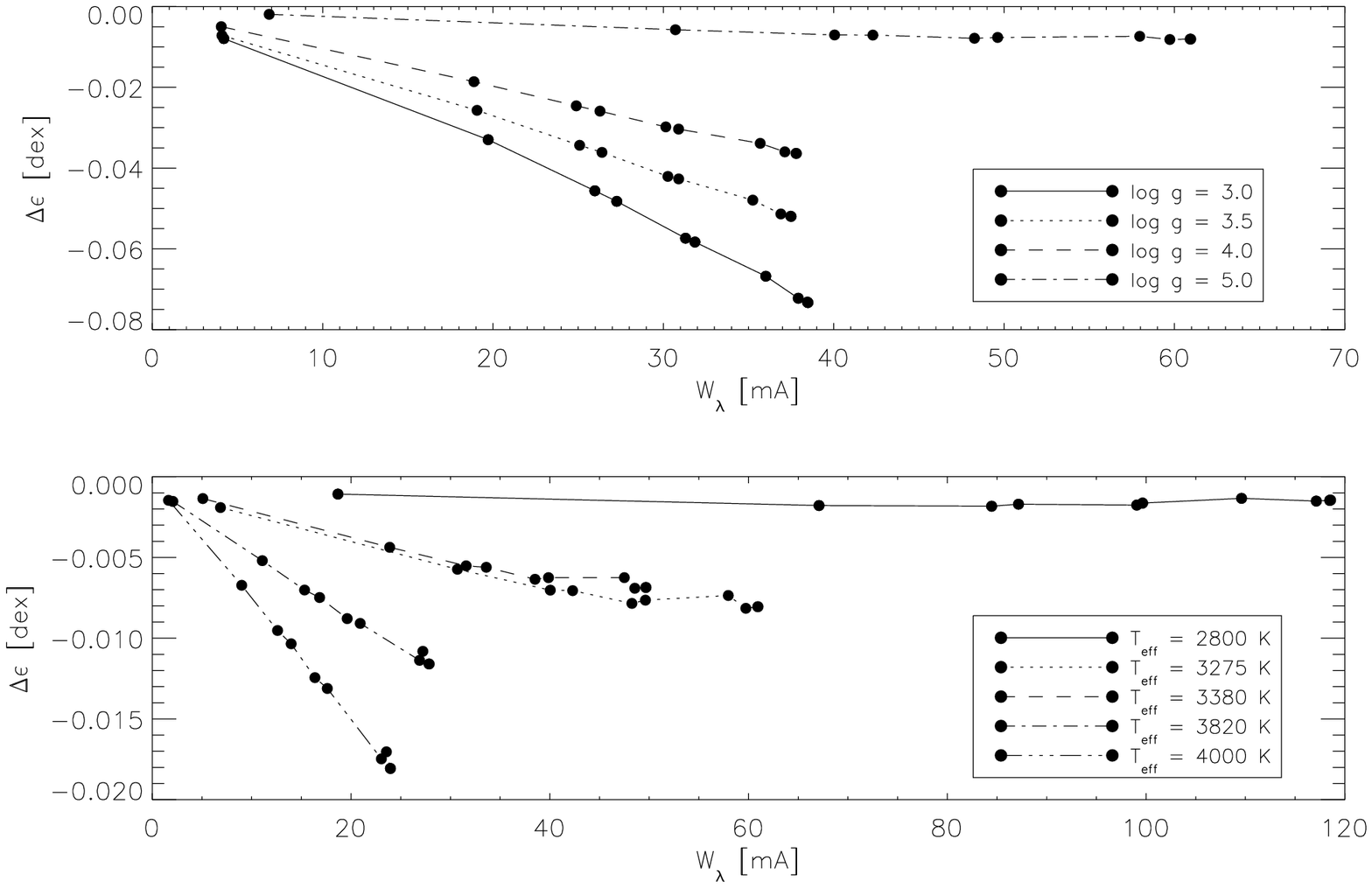}
 \caption{3D -- \mD~corrections to \element[][]{FeH} abundances derived from
  different \element[][]{FeH} lines with varying quantities see Tab.~\ref{tab2}.  
  Upper panel: The models are located at a $\teff$ around
  $3300K$ and different $\log{g}$ values of 3.0 ($\teff~=~3240$\,K), 3.5
  ($\teff~=~3270$\,K), 4.0 ($\teff~=~3315$\,K) and 5.0 ($\teff~=~3275$\,K)
  [cgs].
  Lower panel: The models are located at $\teff$ values of $2800$\,K, $3275$\,K,
  $3380$\,K, $3820$\,K and $4000$\,K and a $\log{g}$ value of 5.0, except the one
  with $\teff~=~3820$\,K ($\log{g}~=~4.9$), and the one with $\teff~=~4000$\,K
  ($\log{g}~=~4.5$)\,[cgs].}
 \label{abucorr}
\end{figure}

\subsubsection{Equivalent Width $W_{\lambda}$}
In the $\teff$-series, $W_{\lambda}$ (see Fig.~\ref{teffdata}
upper panel) decreases with increasing $\teff$. At higher $\teff$ the number
of \element[][]{FeH} molecules decrease due to dissociation and hence
$W_{\lambda}$.
This can be seen in the 3D lines as well as in the \mD-lines.
At $\teff=2800$\,K, the influence of v.d.Waals broadening in the 3D- and
\mD-lines becomes clearly visible in the line profile due to saturation of the \element[][]{FeH}
lines. 
The ten different \element[][]{FeH} lines all behave in a similar
manner. The only difference is the absolute value of $W_{\lambda}$, which depends
on the $\log{gf}$-value and excitation potential $E_l$ of each line.

In the $\log{g}$-series, the $W_{\lambda}$ (see Fig.~\ref{loggdata}
upper panel) increases with increasing $\log{g}$. The change in concentration
of \element[][]{FeH} with lower $\log{g}$, which results in smaller $W_{\lambda}$, 
depends on the changing pressure and density stratification. The difference
between 3D- and \mD-lines at small $\log{g}$-values stems from the
broadening by micro-turbulent velocities and vanishes to higher $\log{g}$ values. This
time the \element[][]{FeH} lines are only mildly saturated, but the velocity
fields in the RHD models (see Sec.3) are strong enough to affect the
$W_{\lambda}$ as well.
Like in the $\teff$-series, the ten different lines show
no significant variations in their behavior. They only vary in the amount of
$W_{\lambda}$ due to different $\log{gf}$-values.     

Since the differences in $W_{\lambda}$ are very small, one can expect that the
3D correction to the \element[][]{FeH} abundance is very small too. We derive
abundance corrections from comparison between 3D and \mD~curve of growths for
each set of lines on the different model atmospheres. The results are plotted
in Fig.~\ref{abucorr}. In this case the correction to the different $\teff$ of
the $\log{g}$ models is not applied. The 3D-\mD~abundance correction is
between $-0.001$\,dex for the coolest high $\log{g}$ model and $-0.07$\,dex for
the $\log{g}=3.0$ model. In all cases the abundance correction is negative
which mean that the 3D lines appear stronger due to the enhanced opacity which
becomes larger due to the micro-turbulent velocity.     
\subsubsection{FWHM}
The dependence of the line width (measured as the width of the line
at their half maximum (FWHM)) on $\teff$ is shown in the middle
panel of Fig.~\ref{teffdata}. At low $\teff$, one can see that the FWHM of the
3D- and \mD-\element[][]{FeH} lines decreases with increasing $\teff$.
The v.d.Waals broadening loses influence and also the dissociation of
\element[][]{FeH} molecules leads to smaller and
narrower lines. After $\teff$ around $3380$\,K, the FWHM of the 3D lines reaches a
flat minimum and starts to become larger again to higher $\teff$. This rise in the line
width is probably related to the rising velocity fields in the RHD models, since
thermal broadening takes place in both, 3D- and \mD-lines and the
latter still decrease. The rise of the velocity in the models with
a $\teff$ of $3380$\,K and $4000$\,K could also be due to the slightly lower
surface gravity in these models, but we think that the main influence stems
from the higher temperatures.  The \mD-lines decrease monotonically
with increasing $\teff$ and reflects the behavior of $W_{\lambda}$. The
difference in FWHM between 3D- and \mD-lines is very small at
$\teff=2800$\,K and increases with increasing $\teff$ to $\sim~0.8$\,km\,s$^{-1}$ at the
highest $\teff$. We have seen in Sec.3.2 that this can be explained with the micro-
and macro-turbulence description. The offset between the FWHM of the ten lines
is due to their different $\log{gf}$-values i.e. large $\log{gf}$-values
results in large FWHM. But since we are interested in broadening from the
velocity fields, we have to take into account that lines with small $\log{gf}$-values,
i.e. weak lines, formed deeper in the atmosphere where the convective motions
are stronger. These lines are more broadened from the hydrodynamical velocity
fields and hence more widened. We can see in
Fig.~\ref{teffdata} that the difference in FWHM between the ten 3D lines becomes
smaller to high $\teff$. This is also valid for the lines in the
$\log{g}$-series where the difference in FWHM between the ten 3D lines
becomes smaller to small $\log{g}$-values.

In the $\log{g}$-series, the dependence of FWHM (Fig.~\ref{loggdata} middle
panel) is very different for 3D- and \mD-lines. In the 3D case the FWHM
stays almost constant with decreasing surface gravity between $\log{g}$ of $5.0$
and $4.0$ for most lines. This is due to the
smaller amount 
of v.d.Waals broadening, which loses its influence due to lower pressure 
in models with smaller surface gravity. This is compared from the broadening
due to the rising velocity fields. With $\log{g}$ smaller than
$4.0$, the width starts to increase for all lines. This increase of line width in the
3D case is a consequence of the hydrodynamic velocity fields which increase 
strongly with decreasing $\log{g}$. 
In the \mD\ case, without the velocity fields, the FWHM decreases with
decreasing surface gravity and reflects the behavior of the $W_{\lambda}$. The
difference between 3D- and \mD-lines reach its maximal value at
$\log{g}=3.0$ [cgs] and is around $1.3$\,km\,s$^{-1}$. This is on the order of
the velocity fields in the RHD models (see Fig.~\ref{loggabsvelos}). One could
fit 1D spectral synthesis \element[][]{FeH} lines to observed ones (with known
$\teff$) with the micro- and macro-turbulence description (see Sec.~3.2) and it
will be possible with the obtained velocities to determine a surface gravity
with the help of Fig.~\ref{loggvelos}.

We did not see any significant different behavior between
the ten \element[][]{FeH} lines. Neither in the $\teff$-series nor in the
$\log{g}$-series.
 
\subsubsection{Line Depth}
In the bottom panel of Fig.~\ref{teffdata} one can see the dependence of the
line depth on $\teff$. The line depth increases with decreasing $\teff$, and
almost all \element[][]{FeH} lines, except the one with the lowest $\log{gf}$ values
are saturated at $\teff=2800$\,K. The difference in line depth between 3D and
\mD-lines is changing in the $\teff$-interval. At high $\teff$, the line
depth of the \mD-lines is deeper than that of the 3D lines. At $\teff=2800$\,K,
this difference almost vanishes. 
This behavior is due
to the saturation of the \element[][]{FeH} lines at low $\teff$ because both,
the 3D- and \mD-lines reach their maximal depth. The decrease
of the line depth with increasing $\teff$ is due to the dissociation of
the \element[][]{FeH} molecules at higher temperatures.

The line depth of the $\log{g}$-series is shown in the bottom panel of
Fig.~\ref{loggdata}. At low $\log{g}$, the line depths of the 3D and \mD-lines increase
almost linearly with increasing $\log{g}$. The 3D lines increase with a strong
slope and the \mD-lines with a weaker slope. The difference in line
depth between 3D- and \mD-models is maximal at $\log{g}=3.0$ [cgs] and
vanishes almost at $\log{g}=5.0$ [cgs]. It is consistent with the velocity
fields present in the atmospheres of the RHD models broadening the lines
and lower the line strength of the 3D lines. The \mD-lines reflect the
decreasing number of \element[][]{FeH} molecules with decreasing $\log{g}$ due
to the lower pressures.

\section{Summary and Conclusion}
We investigated a set of M-star models with $\teff=2500$\,K - $4000$\,K and
$\log{g}=3.0$ -- $5.0$ [cgs]. For these models, the 3D hydrodynamic radiative
transfer code \texttt{CO$^5$BOLD} was used. The horizontal and vertical
velocity fields in the 3D models were described with a binning method. The
convective turn-over point is clearly visible in the atmospheric velocity
dispersion structure. To investigate the influence of these velocity fields on
spectral line shapes, a description for geometrical projection and
limb-darkening effects was applied. With the use of contribution
functions, we took only these parts in the
atmosphere into account where the lines were formed. The resulting velocity
dispersions range from $400$ -- $1600$\,m\,s$^{-1}$ 
with decreasing $\log{g}$ and with increasing $\teff$ from $200$ --
$1400$\,m\,s$^{-1}$. These values agree well with velocities deduced from line shapes.
We expressed the hydrodynamical velocity fields of the 3D models in terms of
the classical micro- and macro-turbulent 
velocities. With this description and the obtained micro- and macro-turbulent
velocities, it is possible to reproduce 3D spectral lines on 1D atmosphere
models very accurately, hence time consuming 3D treatment of \element[][]{FeH}
molecular lines in the regime of cool stars is not necessary for line profile
analysis. A comparison of 
our velocities with a set of velocities determined from observations with
spectral fitting methods showed that the macro-turbulent velocities agree, but
the micro-turbulent velocities are a factor of two or three smaller than the ones
determined from observations.

A line shift due to the larger up-flowing
area in the convection zone was investigated too. It is on the order of a few
m/s up to $50$\,m\,s$^{-1}$ for a very low gravity model. The time dependent
jitter in line positions is only about m/s and would be reduced to \,mm\,s$^{-1}$ in a
real star, due to high number of contributing elements.

In order to use
\element[][]{FeH} molecular lines for investigations of spectroscopic/physical
properties in cool 
stars (e.g. Zeeman- or rotational broadening), we explored the behavior in a
set of lines on
$\log{g}$ and $\teff$. We investigated ten \element[][]{FeH} lines between
9950\AA\ and 9990\AA\ on our models with the
spectral synthesis code \texttt{LINFOR3D}. \element[][]{FeH}
lines react on different effective temperatures as expected due to
the change in chemical composition and pressure. The lines showed also a weak
dependence on surface gravity due to changing densities and pressure. The
broadening from velocity fields in the 3D models of the $\log{g}$ series is
very strong, but for the $\teff$ series the broadening from velocity fields is
almost covered by v.d.Waals broadening. The difference in line width for hot
models is up to $0.5$\,km\,s$^{-1}$ and for low gravity models around $1$\,km\,s$^{-1}$.  That
means for the 1D spectral synthesis, that one has to include correct micro-
and macro-turbulent velocities for small surface gravities or hot
$\teff$. Due
to the fact that the FWHM $\log{g}$ dependence of \element[][]{FeH} lines goes
in the opposite direction as the $\log{g}$ dependence of the velocity fields,
the \element[][]{FeH} lines become a great mean to measure surface gravities
in cool stars. Because the velocity fields scale with $\log{g}$ and it should
be easily possible to detect them.  

\element[][]{FeH} lines with different quantum numbers do not
show significant differences for both, $\log{g}$- and $\teff$-series. That
means the broadening of the lines does not depend on $J$, $\Omega$, or the
branch. Furthermore lines with weak magnetic sensitivity behave just like
lines with strong magnetic sensitivity. All lines are broadened in the same way
by thermal and hydrodynamical motions. Only the transition probability
expressed in the $\log{gf}$ value influences the behavior of the
lines. The line with the lowest $gf$-values did not saturate at low $\teff$,
but in general they are similar to the other \element[][]{FeH} lines.

It is possible to treat the \element[][]{FeH} molecular lines with different
quantum numbers as
homogenous in the absence of magnetic fields. That allows to use
\element[][]{FeH} lines to measure magnetic fields
\citep{2006ApJ...644..497R,2007ApJ...656.1121R}. 
Hence we conclude, that these lines also are an appropriate mean to measure
magnetic field strength in M-type stars.

\section*{Appendix}
\begin{figure}[!h]
\centering
  \includegraphics[width=0.5\textwidth,bb=15 30 580 410]{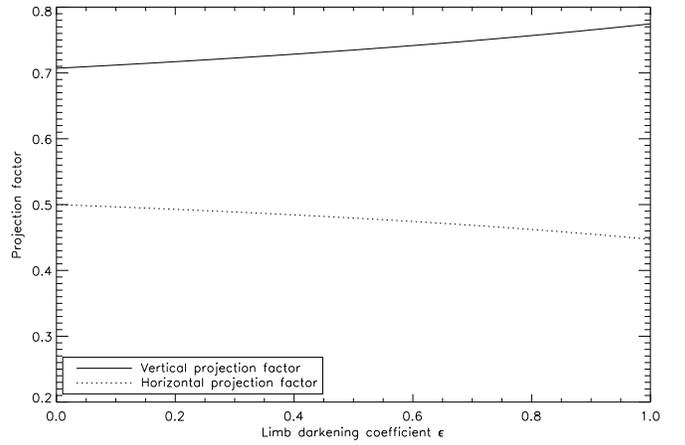}
 \caption{Projection factors (vertical solid line and horizontal dotted line)
 as a function of the limb darkening coefficient $\epsilon$.}
 \label{epsilon}
\end{figure}
To compare the hydrodynamical velocity dispersion in 3D-\texttt{CO$^5$BOLD}
models with the spectroscopic quantities micro- and macro-turbulent
velocities, we assume a simple geometrical model and try to resample the
broadening of absorption lines. An intensity beam ``sees'' the velocity field
under a certain angle and the spectral line is broadened by the projection of
these velocities. In this sense we project the geometrical velocity components
on a line of sight under a certain angle and integrate over all angles in a
half sphere. We also take a linear limb-darkening law into account. We assume
a velocity function $f(v)$ in velocity space and an intensity line profile 
\bq
I(v)=V(v)\cdot I_c^0\cdot L(\theta),
\eq
where $V(v)$ is a line profile function, $I_c^0$
the continuum intensity at the center of the disk and $L(\theta)$ a
limb-darkening law. The integrated flux of the intensity line profile broadened by the
velocity function is then given by their convolution and disk integration, i.e.
\bq
F=\oint f(v) \ast I(v) \cos{\theta}~d\omega,
\eq 
with $d\omega=\sin{\theta}~d\theta~d\phi,~\phi \in [0,2\pi],~\theta \in
[0,\pi/2]$. 
If we assume that $V(v)$ does not vary with different positions on the disk, 
it can be factored out \citep{2008oasp.book.....G} and
\bq
F=V(v)\ast I_c^0\cdot W(v),
\eq
with 
\bq 
W(v)=\oint f(v)\cdot L(\theta) \cos{\theta}~ d\omega.
\label{W}
\eq 
With Eq.~\ref{W} we are left with a flux-like expression for the velocity function.
The dispersion of the velocity function is given by 
\bq
\sigma^2=\langle f(v)^2\rangle-\langle f(v)\rangle^2 
\eq
and we can write for the projected velocity dispersion $\Sigma$:
\bq
\Sigma^2=\frac{1}{N}\int_0^{2\pi}\int_0^{\pi/2}(\langle(\vec{f(v)}\vec{e})^2\rangle-
\langle\vec{f(v)}\vec{e}\rangle^2) \cdot L(\theta)~\cos{\theta}~d\omega
\label{bigint}
\eq
where $\vec{f(v)}=\left(\begin{array}{c} f_x(v) \\ f_y(v) \\ f_z(v)
\end{array} \right)$ is the velocity vector containing all velocities in a
model cube, $\vec{e}=\left(\begin{array}{c} \cos{\phi} \sin{\theta} \\
  \sin{\phi} \sin{\theta} \\ \cos{\theta} \end{array} \right)$ is the basis
vector in spherical coordinates, and
  $L(\theta)=1-\epsilon+\epsilon\cdot\cos(\theta)$ a linear limb-darkening law
with the limb-darkening coefficient $\epsilon$. N is a normalization factor
  $N=\int_0^{2\pi}{d\phi}\int_0^{\pi/2}{L(\theta)~\cos{\theta}~\sin{\theta}~d\theta}=\pi(1-\frac{\epsilon}{3})$.
We included the limb-darkening in the normalization because, in opposite to
the flux, the velocity dispersion for an isotropic velocity field has to be
conserved. The average is taken over the velocities and does not affect the
angle dependent parts.  Basically we have to integrate:
\bq
\langle(\vec{f}\vec{e})^2\rangle-\langle\vec{f}\vec{e}\rangle^2=
(\langle f_x^2\rangle-\langle f_x\rangle^2)\cos{\phi}^2\sin{\theta}^2+\nonumber\\
(\langle f_y^2\rangle-\langle f_y\rangle^2)\sin{\phi}^2\sin{\theta}^2+
(\langle f_z^2\rangle-\langle f_z\rangle^2)\cos{\theta}^2+\nonumber\\
2(\langle f_x f_y\rangle-\langle f_x\rangle \langle f_y\rangle)\cos{\phi}\sin{\phi}\sin{\theta}^2+\nonumber\\
2(\langle f_x f_z\rangle-\langle f_x\rangle \langle f_z\rangle)\cos{\phi}\sin{\theta}\cos{\theta}~~+\nonumber\\
2(\langle f_y f_z\rangle-\langle f_y\rangle \langle f_z\rangle)\sin{\phi}\sin{\theta}\cos{\theta}.~~~~~~
\eq
If we want to compute the average of this quantity, we need the mean and the
squared mean of the quantities, but not the combinations of the velocity components,
because these products vanish in the integration over the half sphere due to
their angle dependent coefficients. After the integration of Eq.\ref{bigint},
$\Sigma^2$ becomes
\bq
\Sigma^2=\frac{(\langle f_x^2 \rangle-\langle f_x \rangle^2 + \langle
  f_y^2\rangle-\langle f_y
  \rangle^2)(7\epsilon-15)}{20(\epsilon-3)}+\nonumber\\ 
\frac{(\langle f_z^2  \rangle-\langle f_z \rangle^2)  
  (6\epsilon-30)}{20(\epsilon-3)}.
\label{sigmapro}
\eq
We can see, that for an isotropic velocity field $f_x(v)=f_y(v)=f_z(v)=f(v)$
follows that $\sigma^2=\langle f(v)^2\rangle-\langle f(v)\rangle^2$ and does
not depend on $\epsilon$ or the geometrical projection anymore. Setting
$\epsilon=0$, it then follows from Eq.~\ref{sigmapro} that
$\Sigma_{x,y}^2=\frac{1}{4}\sigma_{x,y}^2$ and
$\Sigma_z^2=\frac{1}{2}\sigma_z^2$ due to geometrical effects.

The
projection factors for the vertical
$\left(\frac{6\epsilon-30}{20(\epsilon-3)}\right)$ and horizontal
$\left(\frac{7\epsilon-15}{20(\epsilon-3)}\right)$ component are plotted as a
function of the limb-darkening coefficient $\epsilon$ in
Fig.~\ref{epsilon}. They vary only about $5\%$ from no darkening to a full
darkened disk. The reduction for the vertical velocity is about $30\%$ and for
the horizontal components $50\%$.

For the sake of completeness, we obtain in a similar
way the mean velocities in three spatial directions $\langle f_{x,y,z}(v)
\rangle$ which are given by
\bq
\langle f_{x,y,z}(v)\rangle_{projected}=\frac{1}{N}\int_0^{2\pi}\int_0^{\pi/2}\langle \vec{f_{x,y,z}(v)}\vec{e}\rangle
\cdot L(\theta)~\cos{\theta}~d\omega.
\label{smallint}
\eq
The horizontal velocities vanish due to projection but there is still a
vertical component which is reduced to geometrical and limb-darkening effects.
\bq
\langle f_{x}(v)\rangle_{projected}=\langle f_{y}(v)\rangle_{projected}=0, \nonumber\\
\langle f_{z}(v)\rangle_{projected}=\frac{(\epsilon-4)}{2(\epsilon-3)}\langle f_{z}(v)\rangle.~~~~~ 
\eq


\begin{acknowledgements}
SW would like to acknowledge the support from the DFG Research Training
Group GrK - 1351 ``Extrasolar Planets and their host stars''.\\
AR acknowledges research funding from the DFG under an Emmy Noether
Fellowship (RE 1664/4- 1).\\
HGL acknowledges financial support from EU contract MEXT-CT-2004-014265
(CIFIST). We thank Derek Homeier for providing us with the opacity tables.
\end{acknowledgements}

\bibliographystyle{aa}
\bibliography{wende2009_arxiv}

\begin{thebibliography}{36}
\expandafter\ifx\csname natexlab\endcsname\relax\def\natexlab#1{#1}\fi

\bibitem[{{Asplund} {et~al.}(2005){Asplund}, {Grevesse}, \&
  {Sauval}}]{2005ASPC..336...25A}
{Asplund}, M., {Grevesse}, N., \& {Sauval}, A.~J. 2005, in Astronomical Society
  of the Pacific Conference Series, Vol. 336, Cosmic Abundances as Records of
  Stellar Evolution and Nucleosynthesis, ed. T.~G. {Barnes}, III \& F.~N.
  {Bash}, 25--+

\bibitem[{{Barklem} {et~al.}(2000){Barklem}, {Piskunov}, \&
  {O'Mara}}]{2000A&AS..142..467B}
{Barklem}, P.~S., {Piskunov}, N., \& {O'Mara}, B.~J. 2000, \aaps, 142, 467

\bibitem[{{Baschek} {et~al.}(1966){Baschek}, {Holweger}, \&
  {Traving}}]{Bascheck1966}
{Baschek}, B., {Holweger}, H., \& {Traving}, G. 1966, Astronomische Abhandlung
  der Hamburger Sternwarte, 8, 26

\bibitem[{{Bean}(2007)}]{2007PhDT........18B}
{Bean}, J.~L. 2007, PhD thesis, The University of Texas at Austin

\bibitem[{{Bean} {et~al.}(2006{\natexlab{a}}){Bean}, {Benedict}, \&
  {Endl}}]{2006ApJ...653L..65B}
{Bean}, J.~L., {Benedict}, G.~F., \& {Endl}, M. 2006{\natexlab{a}}, \apjl, 653,
  L65

\bibitem[{{Bean} {et~al.}(2006{\natexlab{b}}){Bean}, {Sneden}, {Hauschildt},
  {Johns-Krull}, \& {Benedict}}]{2006ApJ...652.1604B}
{Bean}, J.~L., {Sneden}, C., {Hauschildt}, P.~H., {Johns-Krull}, C.~M., \&
  {Benedict}, G.~F. 2006{\natexlab{b}}, \apj, 652, 1604

\bibitem[{{Dravins}(1982)}]{1982ARA&A..20...61D}
{Dravins}, D. 1982, \araa, 20, 61

\bibitem[{{Dulick} {et~al.}(2003){Dulick}, {Bauschlicher}, {Burrows}, {Sharp},
  {Ram}, \& {Bernath}}]{2003ApJ...594..651D}
{Dulick}, M., {Bauschlicher}, Jr., C.~W., {Burrows}, A., {et~al.} 2003, APJ,
  594, 651

\bibitem[{{Ferguson} {et~al.}(2005){Ferguson}, {Alexander}, {Allard}, {Barman},
  {Bodnarik}, {Hauschildt}, {Heffner-Wong}, \& {Tamanai}}]{jwfOpac05}
{Ferguson}, J.~W., {Alexander}, D.~R., {Allard}, F., {et~al.} 2005, APJ, 623,
  585

\bibitem[{{Freytag} {et~al.}(2009){Freytag}, {Allard}, {Ludwig}, {Homeier}, \&
  {Steffen}}]{bdCO5BOLD}
{Freytag}, B., {Allard}, F., {Ludwig}, H., {Homeier}, D., \& {Steffen}, M.
  2009, A\&A, in prep.

\bibitem[{{Freytag} {et~al.}(2002){Freytag}, {Steffen}, \&
  {Dorch}}]{2002AN....323..213F}
{Freytag}, B., {Steffen}, M., \& {Dorch}, B. 2002, Astronomische Nachrichten,
  323, 213

\bibitem[{{Gray}(1975)}]{1975ApJ...202..148G}
{Gray}, D.~F. 1975, \apj, 202, 148

\bibitem[{{Gray}(1977)}]{1977ApJ...218..530G}
{Gray}, D.~F. 1977, \apj, 218, 530

\bibitem[{{Gray}(2008)}]{2008oasp.book.....G}
{Gray}, D.~F. 2008, {The Observation and Analysis of Stellar Photospheres} (The
  Observation and Analysis of Stellar Photospheres, by D.F.~Gray.~ Cambridge:
  Cambridge University Press, 2008.)

\bibitem[{{Hauschildt} \& {Baron}(1999)}]{1999JCoAM.109...41H}
{Hauschildt}, P.~H. \& {Baron}, E. 1999, Journal of Computational and Applied
  Mathematics, 109, 41

\bibitem[{{Kurucz}(1970)}]{1970SAOSR.309.....K}
{Kurucz}, R.~L. 1970, SAO Special Report, 309

\bibitem[{{Ludwig}(1992)}]{1992HGLPhDT}
{Ludwig}, H.~G. 1992, PhD thesis, University of Kiel

\bibitem[{{Ludwig} {et~al.}(2002){Ludwig}, {Allard}, \&
  {Hauschildt}}]{2002A&A...395...99L}
{Ludwig}, H.-G., {Allard}, F., \& {Hauschildt}, P.~H. 2002, A\&A, 395, 99

\bibitem[{{Ludwig} {et~al.}(2006){Ludwig}, {Allard}, \&
  {Hauschildt}}]{2006A&A...459..599L}
{Ludwig}, H.-G., {Allard}, F., \& {Hauschildt}, P.~H. 2006, A\&A, 459, 599

\bibitem[{{Ludwig} {et~al.}(1994){Ludwig}, {Jordan}, \&
  {Steffen}}]{1994A&A...284..105L}
{Ludwig}, H.-G., {Jordan}, S., \& {Steffen}, M. 1994, \aap, 284, 105

\bibitem[{{Magain}(1986)}]{1986A&A...163..135M}
{Magain}, P. 1986, \aap, 163, 135

\bibitem[{{Nordh} {et~al.}(1977){Nordh}, {Lindgren}, \&
  {Wing}}]{1977A&A....56....1N}
{Nordh}, H.~L., {Lindgren}, B., \& {Wing}, R.~F. 1977, A\&A, 56, 1

\bibitem[{{Nordlund}(1982)}]{1982A&A...107....1N}
{Nordlund}, A. 1982, \aap, 107, 1

\bibitem[{{Palla} \& {Baraffe}(2005)}]{2005A&A...432L..57P}
{Palla}, F. \& {Baraffe}, I. 2005, \aap, 432, L57

\bibitem[{{Reiners}(2007)}]{2007A&A...467..259R}
{Reiners}, A. 2007, A\&A, 467, 259

\bibitem[{{Reiners} \& {Basri}(2006)}]{2006ApJ...644..497R}
{Reiners}, A. \& {Basri}, G. 2006, ApJ, 644, 497

\bibitem[{{Reiners} \& {Basri}(2007)}]{2007ApJ...656.1121R}
{Reiners}, A. \& {Basri}, G. 2007, ApJ, 656, 1121

\bibitem[{{Sauval} \& {Tatum}(1984)}]{1984ApJS...56..193S}
{Sauval}, A.~J. \& {Tatum}, J.~B. 1984, ApJS, 56, 193

\bibitem[{{Schweitzer} {et~al.}(1996){Schweitzer}, {Hauschildt}, {Allard}, \&
  {Basri}}]{1996MNRAS.283..821S}
{Schweitzer}, A., {Hauschildt}, P.~H., {Allard}, F., \& {Basri}, G. 1996,
  MNRAS, 283, 821

\bibitem[{{Sneden}(1973)}]{1973PhDT.......180S}
{Sneden}, C.~A. 1973, PhD thesis, AA(THE UNIVERSITY OF TEXAS AT AUSTIN.)

\bibitem[{{Steffen} {et~al.}(1995){Steffen}, {Ludwig}, \&
  {Freytag}}]{1995A&A...300..473S}
{Steffen}, M., {Ludwig}, H.-G., \& {Freytag}, B. 1995, \aap, 300, 473

\bibitem[{{V{\"o}gler} {et~al.}(2004){V{\"o}gler}, {Bruls}, \&
  {Sch{\"u}ssler}}]{2004A&A...421..741V}
{V{\"o}gler}, A., {Bruls}, J.~H.~M.~J., \& {Sch{\"u}ssler}, M. 2004, \aap, 421,
  741

\bibitem[{{Wedemeyer} {et~al.}(2004){Wedemeyer}, {Freytag}, {Steffen},
  {Ludwig}, \& {Holweger}}]{2004A&A...414.1121W}
{Wedemeyer}, S., {Freytag}, B., {Steffen}, M., {Ludwig}, H.-G., \& {Holweger},
  H. 2004, \aap, 414, 1121

\bibitem[{{Wilkinson}(1963)}]{1963ApJ...138..778W}
{Wilkinson}, P.~G. 1963, \apj, 138, 778

\bibitem[{{Wing}(1972)}]{1972saim.conf..123W}
{Wing}, R.~F. 1972, in Les Spectres des Astres dans l'Infrarouge et les
  Microondes, 123--140

\bibitem[{{Wing} \& {Ford}(1969)}]{1969PASP...81..527W}
{Wing}, R.~F. \& {Ford}, W.~K.~J. 1969, PASP, 81, 527

\end{thebibliography}

\end{document}